\documentclass[reprint,amsfonts, amssymb, amsmath,pra,nofootinbib, superscriptaddress, twocolumn,longbibliography,aps]{revtex4-2}
\usepackage{float}
\makeatletter
\let\newfloat\newfloat@ltx
\makeatother
\usepackage[utf8]{inputenc}
\usepackage[english]{babel}
\usepackage{graphics}
\usepackage{selinput}
\usepackage[normalem]{ulem}
\usepackage[shortlabels]{enumitem}
\usepackage{physics}
\usepackage{braket}
\usepackage{amsthm}
\usepackage{mathtools}
\usepackage[caption=false,font=small]{subfig}
\usepackage{svg}
\svgsetup{inkscapelatex=true}
\usepackage{booktabs}
\definecolor{darkgreen}{rgb}{0,0.5,0}
\usepackage{pifont}
\usepackage{xcolor}
\usepackage{graphicx}
\usepackage{adjustbox}
\usepackage{placeins}
\usepackage[T1]{fontenc}
\usepackage{lipsum}
\usepackage{csquotes}
\usepackage{bm}
\usepackage{bbm}

\usepackage{cellspace}
\usepackage[linesnumbered,ruled,vlined]{algorithm2e}
\SetKwInput{kwInit}{Init}

\usepackage[makeroom]{cancel}
\usepackage[toc,page]{appendix}
\usepackage[colorlinks=true,allcolors=blue]{hyperref}

\usepackage{tikz}
\tikzset{every picture/.style=remember picture}



\definecolor{pa}{rgb}{0,.4,1}

\makeatletter
\newif\ifeqcontrib@this
\newif\ifeqcontrib@any
\eqcontrib@thisfalse
\eqcontrib@anyfalse
\newcommand{\eqcontrib}{\global\eqcontrib@thistrue\global\eqcontrib@anytrue}
\newcommand{\eqcontribmark}{\textsuperscript{\ensuremath{,\dagger\dagger}}}

\newcommand{\eqcontrib@maybe}{%
  \ifeqcontrib@this
    \global\eqcontrib@thisfalse
    \eqcontribmark
  \fi
}

\usepackage[most]{tcolorbox}

\newtcolorbox{myresult}[1]{
  colback=blue!5!white,    
  colframe=blue!75!black,  
  fonttitle=\bfseries,     
  title=#1,                
  arc=4mm,                 
  outer arc=1mm
}
\newcommand{\printEqContrib}{%
  \ifeqcontrib@any
    \begingroup
      \renewcommand{\thefootnote}{\ensuremath{\dagger\dagger}}%
      \footnotetext{These authors contributed equally to this work.}%
    \endgroup
  \fi
}

\def\doauthor#1#2#3{%
  \ignorespaces#1\unskip\@listcomma
  \begingroup
    #3%
  \@if@empty{#2}{\endgroup{}{}}{\endgroup{\comma@space}{}\frontmatter@footnote{#2}}%
  \eqcontrib@maybe
  \space \@listand
}%
\makeatother

\begin{document}

\title{Dynamical, thermal, and ground-state multiparameter quantum metrology: Fundamental limits and their attainability in magnetometry 
}
\author{Riccardo Pedroni}
\thanks{riccardo.pedroni@inria.fr}
\affiliation{Física Teòrica: Informació i Fenòmens Quàntics, Departament de Física, Universitat Autònoma de Barcelona, 08193 Bellaterra (Barcelona), Spain}
\affiliation{Dipartimento di Fisica e Astronomia, Università di Bologna, via Irnerio 46, I-40126 Bologna, Italy}
\author{Víctor Izquierdo}
\thanks{victor.izquierdo@uab.cat}
\affiliation{Física Teòrica: Informació i Fenòmens Quàntics, Departament de Física, Universitat Autònoma de Barcelona, 08193 Bellaterra (Barcelona), Spain}
\author{Advay Burte}
\thanks{f20212873@goa.bits-pilani.ac.in}
\affiliation{Department of Physics, BITS Pilani KK Birla Goa Campus, Zuarinagar 403726, Goa, India}
\affiliation{Harish-Chandra Research Institute,  Chhatnag Road, Jhunsi, Prayagraj - 211019, India}
\author{Paolo Abiuso}
\thanks{paolo.abiuso@oeaw.ac.at}
\affiliation{Institute for Quantum Optics and Quantum Information - IQOQI Vienna,
Austrian Academy of Sciences, Boltzmanngasse 3, A-1090 Vienna, Austria}
\author{John Calsamiglia}
\thanks{john.calsamiglia@uab.cat}
\affiliation{Física Teòrica: Informació i Fenòmens Quàntics, Departament de Física, Universitat Autònoma de Barcelona, 08193 Bellaterra (Barcelona), Spain}
\author{Martí Perarnau-Llobet}
\thanks{marti.perarnau@uab.cat}
\affiliation{Física Teòrica: Informació i Fenòmens Quàntics, Departament de Física, Universitat Autònoma de Barcelona, 08193 Bellaterra (Barcelona), Spain}

\date{\today}

\begin{abstract}
We study the simultaneous estimation of parameters in Hamiltonians $H_{\vec{\theta}}=H_{\vec{\theta}}^P+H^C$, with $H^C$ a control term, for unitary dynamics, thermal equilibrium, and ground states. In each case, we bound the quantum Fisher information matrix by the covariance matrix of the parameter generators, extending single-parameter bounds and identifying the conditions for saturation. At thermal equilibrium, a complementary lower bound quantifies the cost of non-commutativity. For vector magnetometry, we derive asymptotically tight bounds and identify many-body control Hamiltonians that approach them in all three settings.
\end{abstract}
\maketitle

\section{Introduction}
Quantum metrology employs quantum systems to enhance parameter estimation beyond classical limits and to investigate the fundamental bounds imposed by quantum mechanics \cite{Paris_2009, Giovannetti_2011, Degen_2017}. While early breakthroughs in the field focused mostly on single-parameter estimation, many quantum sensing tasks require the simultaneous estimation of multiple parameters, such as in quantum imaging, phase-tracking, and vector field sensing.  This has led to a renewed interest in understanding the fundamental limits that quantum mechanics imposes on the simultaneous estimation of several parameters \cite{Szczykulska_2016,Albarelli_2020,Albarelli_2022,pezze_2025}.

The joint sensitivity of a quantum probe to a vector of parameters $\vec{\theta}$ is encoded in the Quantum Fisher Information Matrix (QFIM), $Q$, whose diagonal elements quantify individual sensitivities whereas off-diagonal elements encode statistical correlations \cite{Liu_2019, Gessner_2018, pezze_2025}. In particular, the scalar figure of merit $\text{Tr}[WQ^{-1}]$, where $W$ is a positive-definite weight matrix, bounds the weighted mean square error of any unbiased estimator via the Quantum Cramér--Rao bound. In contrast to the single-parameter case, and due to the inherent incompatibility of optimal measurements in quantum mechanics, this bound is not saturable in general, making the search for optimal estimation strategies in the multiparameter regime highly involved \cite{Albarelli2019Evaluating,DemkowiczDobrzaski2020,xia2023toward_incompatible_quantum_limits,pezze_2025}.

The QFIM naturally depends on how the unknown parameters $\vec{\theta}$ are imprinted into the quantum probe. In a typical metrological setup, this is described by a parameter-dependent Hamiltonian $H_{\vec{\theta}}^P$. To enhance the QFIM, one can introduce a control term $H^C$ under experimental control, yielding a total Hamiltonian of the form $H_{\vec{\theta}}=H_{\vec{\theta}}^{P}+H^{C}$. In this work, we analyze fundamental bounds on $\text{Tr}[WQ^{-1}]$ for three paradigmatic control-Hamiltonian encodings: (i) unitary dynamics, (ii) thermal equilibrium (Gibbs states), and (iii) ground states.

The properties of the QFIM and the resulting precision limits have been thoroughly investigated for these scenarios. For dynamical metrology, the potential of entangled states to surpass classical strategies, together with trade-offs between optimal measurements in multiparameter estimation tasks, has been characterized~\cite{Vaneph2013, Yuan2016, Hou_2020}. Bounds in the presence of noise have also been developed \cite{roy2019fundamental_noisy_multiparameter_bounds, Albarelli_2022, brady2026precisionlimitsmultiparametermarkoviannoise}. For probes in equilibrium, the potential of many-body interactions and criticality has been investigated \cite{Montenegro_2025,mihailescu2025criticalquantumsensingtutorial}, including multiparameter estimation tasks such as the simultaneous estimation of temperature and magnetic fields \cite{Mihailescu_2024, Ullah2026Configuration, ullah2026enhancing, previdi2025oh,previdi2026multi,Mihailescu2025Uncertain, bachain2026multiparameterquantummetrologymolecular}. Notably, the general form of the QFIM for critical probes has been established in \cite{difresco2022multiparameter, campos2007quantum, DIFRESCO2025116913}.

Despite these remarkable advances, the ultimate limits of multiparameter metrology have not yet been established, in the sense of finding the optimal $H^C$ that optimizes the QFI matrix for a given Hamiltonian signal $H_{\vec{\theta}}^P$ and encoding (dynamical, thermal, or ground state). For dynamical metrology, optimizing over $H^C$ is complementary to the standard optimization over the initial state \cite{Puig2025Dynamical, oconnor2026geometricobstructions}; however, for Gibbs state metrology, $H^C$ is the crucial resource enabling enhanced sensitivity \cite{Abiuso_2025}. In the single-parameter case, the problem reduces to maximizing the scalar QFI $\mathcal{F}_\theta$, and fundamental bounds for unitary, thermal equilibrium, and ground-state metrology are known \cite{Boixo_2007, Abiuso_2025}; for thermal states, bounds on the diagonal elements of the QFIM were also derived in~\cite{GarciaPintos2024}. In this work, we generalize these results by deriving general bounds on $\text{Tr}[WQ^{-1}]$, along with the corresponding conditions for saturation. They follow from matrix inequalities between the QFIM and the covariance matrix of the generators; for thermal states, a complementary lower bound on the QFIM in terms of a double commutator shows when they are approached, even if they cannot be saturated exactly.

We then focus on vector magnetometry, a genuinely multiparameter estimation task of crucial interest in quantum metrology, which has experienced significant theoretical \cite{Albarelli2019Evaluating,Hou_2020,Grecki2022,Baamara2023,Hayashi2024,apellaniz2025differential} and experimental advances \cite{Hou_2021, polino_2020, jiang2021multiparameterquantummetrologyusing, meng_2023, rieckmann2026vector,isogawa2026approaching,Clevenson2015,Chipaux2015,Schloss2018,Li2026}. Here, we derive asymptotically tight bounds for the simultaneous estimation of two and three components of weak magnetic fields, and identify unitary, ground-state, and thermal-state configurations that saturate them or, in one case, attain them asymptotically. These bounds complement previous results in multiparameter unitary magnetometry \cite{Baumgratz_2016, Hou_2020,oconnor2026geometricobstructions} and provide guidelines for the implementation of optimal quantum probes in vector magnetometry at thermal equilibrium and for ground states.

\section{Framework}
We consider the estimation of $d$ unknown parameters $\{\theta_i\}_{i=1}^d$ encoded
in a quantum state $\rho_{\vec{\theta}}$, measured by a POVM with outcomes $\vec{x}$
from which estimates $\hat{\theta}_i(\vec{x})$ are built. We work in the local
frequentist setting \cite{Li_2018}, where locally unbiased estimators are
characterized by the covariance matrix
$V_{ij} \coloneqq \mathrm{Cov}(\hat{\theta}_i,\hat{\theta}_j)$ and, as a scalar
figure of merit, by the weighted mean square error (WMSE) $\mathrm{Tr}[WV]$, with
$W>0$ a cost matrix. The quantum Cram\'er--Rao bound (QCRB) $V \geq Q^{-1}$
\cite{holevo1982probabilistic,Liu_2019} then gives
\begin{equation}
    \mathrm{Tr}[WV] \ge \mathrm{Tr}[WQ^{-1}],
    \label{scalarQCRB}
\end{equation}
where the quantum Fisher information matrix (QFIM) is
$Q_{ij} \coloneqq \mathrm{Tr}[\rho_{\vec{\theta}}\{L_i,L_j\}]/2$, with the symmetric
logarithmic derivatives (SLDs) defined by $\partial_i\rho_{\vec{\theta}} =\{L_i,\rho_{\vec{\theta}}\}/2$.

Differently from the one-parameter case \cite{holevo1982probabilistic}, this bound
is not always attainable: allowing collective measurements on many copies of the
probe, it is asymptotically saturable if and only if the weak commutativity
condition (WCC) is satisfied
\cite{Ragy_2016,Kahn_2009,Yamagata_2013,Yang_2019},
\begin{equation}
    \mathrm{Tr}(\rho_{\vec{\theta}} [L_i,L_j]) = 0,
    \quad \forall\, i,j = 1,\ldots,d.
    \label{eq:WCC}
\end{equation}
Even when it is not, an asymptotic precision bounded by $2\,\mathrm{Tr}[WQ^{-1}]$
remains achievable \cite{Albarelli_2020}.

\subsection*{Metrological settings}
For a fixed $H_{\vec{\theta}}^P$, different encoding processes, such as unitary evolution, thermalization, and ground-state preparation, give rise to different families of probe states $\{\rho_{\vec{\theta}}\}_{\vec{\theta}}$. In each of them, we add a term $H^C$ under experimental control, so that the total Hamiltonian is 
\begin{equation*}
    H_{\vec{\theta}}=H_{\vec{\theta}}^P+H^C.
\end{equation*}

In unitary metrology, the quantum probe is prepared in an initial state $\rho_0$ and evolves according to the unitary operator $U_{\vec{\theta}}=\exp(-itH_{\vec{\theta}}/\hbar)
$, so that
\begin{equation}
    \rho_{\vec{\theta}} = U_{\vec{\theta}} \rho_0 U_{\vec{\theta}}^\dagger.
    \label{eq:unitaryevolvedstate}
\end{equation}
In this setting, the WMSE is minimized over both the initial state $\rho_0$ and the control Hamiltonian $H^C$.

In thermal equilibrium metrology, the probe is brought into contact with a thermal bath and thermalizes to the Gibbs state
\begin{equation}
\rho_{\vec{\theta}} =
\frac{e^{-\beta H_{\vec{\theta}}}}
{\operatorname{Tr}(e^{-\beta H_{\vec{\theta}}})},
\label{eq:Gibbsstate}
\end{equation}
where $\beta=(k_B T)^{-1}$, $k_B$ is the Boltzmann constant, and $T$ denotes the temperature. In this case, the WMSE is minimized by optimizing only the control Hamiltonian $H^C$.

In the zero-temperature limit, thermal metrology reduces to ground-state metrology. Assuming a non-degenerate ground state, the probe is then described by the ground state of $H_{\vec{\theta}}$.

\section{General bounds}
We now present general bounds for $\mathrm{Tr}[WQ^{-1}]$ for unitary dynamics, at thermal equilibrium, and for ground states. We assume that all parameters are to be estimated, so that $W > 0$, and restrict the optimization to states for which $Q$ is invertible, since otherwise the probe is insensitive to some parameter combinations and the right-hand side of Eq.~\eqref{scalarQCRB} diverges\footnote{When the QFIM is not invertible, a bound on the precision of the sensitive parameter combinations can be obtained by replacing $Q^{-1}$ with the Moore--Penrose pseudoinverse of $Q$ \cite{Penrose1955, Namkung_2025}.}. We further assume $W$ to be diagonal, $W_{ij} = \delta_{ij} w_i$, so that the error of each parameter is weighted independently\footnote{If $W=ODO^T$ is not diagonal, with $O$ orthogonal and $D$ diagonal, the problem maps onto the estimation of the rotated parameters $\vec{\lambda}=O^T\vec{\theta}$, whose QFIM is $O^TQO$, with diagonal cost matrix $D$.}.
All our bounds follow from a common structure: in each setting, the QFIM is upper bounded, in the matrix sense, by a covariance matrix of the generators of the parameters,
\begin{equation}
    Q\leq\kappa\,\Gamma,\qquad \Gamma_{ij}=\mathrm{Cov}_\rho(G_i,G_j),
    \label{eq:QleqGamma}
\end{equation}
where $\mathrm{Cov}_\rho(A,B)=\frac{1}{2}\braket{\{A,B\}}_\rho-\braket{A}_\rho\braket{B}_\rho$, and the generators $G_i$, the state $\rho$, and the constant $\kappa>0$ depend on the setting. Since matrix inversion reverses the order between positive matrices, and $(\Gamma^{-1})_{ii}\geq1/\Gamma_{ii}$~\cite{Albarelli_2022},
\begin{equation}
    \mathrm{Tr}[WQ^{-1}] \geq \frac{1}{\kappa}\mathrm{Tr}[W\Gamma^{-1}] \geq \frac{1}{\kappa}\sum_{i=1}^d \frac{w_i}{\mathrm{Var}_\rho(G_i)}.
    \label{eq:boundtraceQFIM}
\end{equation}
For diagonal $W$, the first inequality is saturated if and only if $Q=\kappa\Gamma$, and the second one whenever $\Gamma$ is diagonal. Probe-independent bounds then follow from $\mathrm{Var}_\rho(A)\leq\|A\|^2/4$~\cite{Abiuso_2025}, where $\|{A}\| \coloneqq \lambda_\mathrm{max}(A)-\lambda_\mathrm{min}(A)$ denotes the seminorm, with $\lambda_\mathrm{max}(A)$ ($\lambda_\mathrm{min}(A)$) the maximum (minimum) eigenvalue of $A$. Since these maxima need not be reached simultaneously, the resulting bounds are not tight in general (see Appendix~\ref{app:upperboundvariance}).

Let us first focus on the paradigmatic case of \emph{unitary metrology}. Here $G_i=\mathcal{H}_i\coloneqq i (\partial_i U_{\vec{\theta}}^\dagger)U_{\vec{\theta}}$, covariances are evaluated in the initial state $\rho_0$, and $\kappa=4$, with equality in Eq.~\eqref{eq:QleqGamma} for pure $\rho_0$~\cite{Liu_2019}. As shown in Appendix~\ref{app:boundunitary}, the following inequalities hold:
 \begin{align}
    \mathrm{Tr}[WQ^{-1}] & \geq \frac{1}{4}\mathrm{Tr}[W\Gamma^{-1}]
    \geq \min_{\rho_0}\sum_{i=1}^d \frac{w_i}{4\mathrm{Var}_{\rho_0} (\mathcal{H}_i)}\notag \\&\geq\sum_{i=1}^d \frac{w_i}{\| \mathcal{H}_i \|^2}
    \geq \sum_{i=1}^d \frac{w_i\hbar^2}{t^2 \|\partial_i H_{\vec{\theta}}\|^2}. \label{eq:boundunitary}
\end{align}
The first inequality is saturated for pure initial states, the second one requires $\Gamma$ to be diagonal, and the fourth holds, e.g., if $[H_{\vec{\theta}}, \partial_i H_{\vec{\theta}}]=0$ for all $i$.

Let us now focus on \emph{thermal equilibrium metrology}. In Appendix~\ref{app:QFIMthermal} we show that, for a Gibbs state as in Eq.~\eqref{eq:Gibbsstate}, the QFIM can be written as:
\begin{equation*}
    Q_{ij} = \beta^2 \left( \mathrm{Tr}[\, \partial_i H_{\vec{\theta}} \, \mathcal{J}_ {\rho}[\partial_jH_{\vec{\theta}}] \,] - \mathrm{Tr}[\rho_{\vec{\theta}} \partial_i H_{\vec{\theta}}]\mathrm{Tr}[\rho_{\vec{\theta}} \partial_jH_{\vec{\theta}}]\right),
\end{equation*}
where the superoperator $\mathcal{J}_\rho$ is defined as $\mathcal{J}_\rho = \mathbb{J}_{L,\rho} \circ \mathbb{J}^{-1}_{B,\rho} \circ \mathbb{J}_{L,\rho}$, with $\mathbb{J}_{L,\rho}[A] = \int_0^1 ds\, \rho_{\vec{\theta}}^s A \rho_{\vec{\theta}}^{1-s}$ denoting the logarithmic multiplication superoperator, and $\mathbb{J}_{B,\rho}[A] = (\rho_{\vec{\theta}} A + A \rho_{\vec{\theta}})/2$ the Bures multiplication superoperator~\cite{scandi2024quantumfisherinformationdynamical}. Here $G_i=\partial_iH_{\vec{\theta}}$, covariances are evaluated in the Gibbs state, and $\kappa=\beta^2$. Moreover, Eq.~\eqref{eq:QleqGamma} can be complemented by a lower bound (see Appendix~\ref{app:thermalsandwich}),
\begin{equation}
  \beta^2\Gamma-\frac{\beta^3}{6}K\;\leq\;Q\;\leq\;\beta^2\Gamma,
  \label{eq:thermalsandwich}
\end{equation}
where $K_{ij}=\braket{[\partial_iH_{\vec{\theta}},[H_{\vec{\theta}},\partial_jH_{\vec{\theta}}]]}_\rho$ is positive semidefinite. The upper bound was obtained for the diagonal elements in~\cite{GarciaPintos2024,Abiuso_2025} and in matrix form in~\cite{Cao_2026}; the lower bound can be seen as a QFI analogue of the Harris inequality~\cite{Harris1967}, and quantifies the effect of non-commutativity through a double commutator. From Eqs.~\eqref{eq:boundtraceQFIM} and~\eqref{eq:thermalsandwich},
\begin{align}
    \mathrm{Tr}[WQ^{-1}] &\geq \frac{1}{\beta^2}\mathrm{Tr}[W\Gamma^{-1}] \geq \sum_{i=1}^d \frac{w_i}{\beta^2 \mathrm{Var}_{\rho}(\partial_i H_{\vec{\theta}})}\label{eq:boundthermal} \\
    &\geq \min_{H^C }\sum_{i=1}^d \frac{w_i}{\beta^2 \mathrm{Var}_{\rho}(\partial_i H_{\vec{\theta}})}
    \geq \sum_{i=1}^d \frac{4w_i}{\beta^2 \|\partial_i H_{\vec{\theta}}\|^2}, \notag
\end{align}
while, for any probe with $\Gamma>\beta K/6$,
\begin{equation}
    \mathrm{Tr}[WQ^{-1}]\leq\frac{1}{\beta^2}\mathrm{Tr}\Big[W\Big(\Gamma-\frac{\beta}{6}K\Big)^{-1}\Big].
    \label{eq:thermalupper}
\end{equation}
The first inequality in Eq.~\eqref{eq:boundthermal} is saturated if and only if $K=0$ (see Theorem~1 in Appendix~\ref{app:thermalsandwich}), i.e., $[H_{\vec{\theta}},\partial_iH_{\vec{\theta}}]=0$ for all $i$; by the Jacobi identity, $H_{\vec{\theta}}$ then commutes with the whole Lie algebra generated by the $\partial_iH_{\vec{\theta}}$, including elements that are not estimated. Otherwise, Eq.~\eqref{eq:thermalupper} shows that the bound is approached whenever $\beta K\ll\Gamma$, i.e., when the non-commutativity is weak on the thermal energy scale. Exact commutation holds, for instance, if $H_{\vec{\theta}}\propto\mathbb{I}$ at the working point, e.g., for $H_{\vec{\theta}}=\sum_i\theta_ih_i$ around $\vec{\theta}\approx0$, whose Gibbs state is maximally mixed.

We now consider \emph{ground-state (GS) metrology}. In the zero-temperature limit, the previous bound becomes trivial, vanishing as $\beta^{-2}$, since degenerate ground states can make the QFI diverge.

In many realistic setups, however, the ground state is non-degenerate and the minimum spectral gap is a finite quantity $\Delta$. In this case, as shown in Appendix~\ref{app:QFIMGS} (see also Refs.~\cite{campos2007quantum,zanardi2007differentialinformationgeometryquantumphase}), the QFIM can be written as
\begin{equation}
    Q_{ij} = \sum_{k>0} \frac{4\,\mathrm{Re}\big[(\partial_i H_{\vec{\theta}})_{0k}(\partial_j H_{\vec{\theta}})_{k0}\big]}{(E_k - E_0)^2} + \mathcal{O}(\beta^2e^{-\beta\Delta}),
    \label{eq:QFIMGSformula}
\end{equation}
where $H_{\vec{\theta}} = \sum_k E_k \ket{k}\bra{k}$ and $A_{kl}\coloneqq\bra{k}A\ket{l}$. Since $E_k-E_0\geq\Delta$, Eq.~\eqref{eq:QFIMGSformula} implies Eq.~\eqref{eq:QleqGamma} with $G_i=\partial_iH_{\vec{\theta}}$ in the ground state and $\kappa=4/\Delta^2$ (see Appendix~\ref{app:boundGS}), so that
\begin{align}
    \mathrm{Tr}[WQ^{-1}] &\geq \frac{\Delta^2}{4}\mathrm{Tr}[W\Gamma^{-1}]\geq \min_{H^C} \sum_{i=1}^d \frac{w_i \Delta^2}{4\,\mathrm{Var}_\rho(\partial_i H_{\vec{\theta}})}\notag\\ &\geq \sum_{i=1}^d \frac{w_i \Delta^2}{\lVert \partial_i H_{\vec{\theta}} \rVert^2}. 
    \label{eq:boundGS}
\end{align}
The first inequality is saturated if the $\partial_iH_{\vec{\theta}}$ couple the ground state only to the first excited level(s), i.e., $(\partial_iH_{\vec{\theta}})_{0j}=0$ whenever $E_j-E_0\neq\Delta$, and the second if $\Gamma$ is diagonal; finding a control Hamiltonian that satisfies both is highly non-trivial.

\section{Fundamental bounds in vector magnetometry}
We now derive tight bounds in vector magnetometry, namely for a parameter-encoding Hamiltonian of the form
\begin{equation}
    H^P_{\vec{\theta}} = \sum_{i=1}^d \theta_i J_i, \label{eq:hamiltonianmagnetometry}
\end{equation}
where $d=1,2,3$ denotes the dimensionality of the field and \(J_i = \sum_{j=1}^N \sigma_i^{(j)}/2\) is the \(i\)-th component of the total angular momentum. Here, \(N\) is the number of particles in the system, \(\sigma_i^{(j)}\) the Pauli matrices, and we set \(\hbar = 1\). This is a paradigmatic setting in which the variances of the different derivatives
cannot be maximized simultaneously: while a single one can reach
$\mathrm{Var}_\rho(J_i)=\|J_i\|^2/4=N^2/4$, they are jointly constrained by
$\sum_{i=1}^3\mathrm{Var}_\rho(J_i)\leq\braket{\vec{J}^2}_\rho\leq N(N+2)/4$.
We further assume the normalization condition $\sum_{i=1}^d \sqrt{w_i} = d$. Note that a global rescaling of the cost matrix does not affect the optimal configuration. With this normalization, the unweighted case is recovered for $w_i = 1$, corresponding to $W = \mathbb{I}_d$.

Under these assumptions, as shown in Appendix~\ref{app:boundsmagnetometry}, the following inequalities hold respectively for unitary, thermal equilibrium and GS metrology (see also Table~\ref{tab:QFIM_fundamentalbounds_magnetometry}):
\begin{align}
    \mathrm{Tr}[WQ^{-1}]
    &\geq \min_{\rho_0}\sum_{i=1}^d \frac{w_i}{4\,\mathrm{Var}_{\rho_0} (\mathcal{H}_i)}
    \geq \frac{d^2}{t^2 N(N+2)},
    \label{eq:boundmagnetometryunitary}\\
    \mathrm{Tr}[WQ^{-1}]
    &\geq \min_{H^C} \sum_{i=1}^d \frac{w_i}{\beta^2 \mathrm{Var}_{\rho}(J_i)}
    \geq \frac{4d^2}{\beta^2 N(N+2)}, \label{eq:boundmagnetometrythermal}\\
    \mathrm{Tr}[WQ^{-1}]
    &\geq \min_{H^C} \sum_{i=1}^d \frac{w_i \Delta^2}{4\,\mathrm{Var}_{\rho}(J_i)}
    \geq \frac{\Delta^2 d^2}{N(N+2)}.\label{eq:boundmagnetometryGS}
\end{align}

The first inequality in each of the three bounds is saturated under the respective conditions discussed above, i.e., when $Q=\kappa\Gamma$ with $\Gamma$ diagonal. The second inequality, instead, is saturated whenever there exists a state (and, for the thermal and GS bounds, a corresponding control Hamiltonian $H^C$) such that $\mathrm{Var}_\rho(J_i) = \sqrt{w_i}\,N(N+2)/(4d)$ for the $d$ estimated directions, and $\mathrm{Var}_\rho(J_i)=0$ for the remaining ones. For $W=\mathbb{I}_d$, we show below that this is possible in all three settings for $d=3$, and in unitary and GS metrology for $d=2$, so that Eqs.~\eqref{eq:boundmagnetometryunitary}--\eqref{eq:boundmagnetometryGS} are tight in these cases.

The remaining case, two-parameter thermal magnetometry, cannot saturate the bound exactly. Saturation would require $Q=\beta^2\Gamma$, i.e., $[H_{\vec\theta},J_1]=[H_{\vec\theta},J_2]=0$, and hence, by the Jacobi identity, $[H_{\vec\theta},J_3]=0$. The Gibbs state would then be rotationally invariant, with $\mathrm{Var}_\rho(J_3)=\mathrm{Var}_\rho(J_1)\neq0$, whereas saturation requires $\mathrm{Var}_\rho(J_3)=0$. The best commuting probe, $H^C=-MJ^2$, misses the bound by a factor $3/2$ (see Appendix~\ref{app:nogod2}). Nevertheless, Eq.~\eqref{eq:thermalupper} identifies non-commuting probes whose WMSE exceeds the bound only by a factor $1+\mathcal{O}(N^{-1})$, so that the attainable precision has the same large-$N$ form, prefactor included (see below).
\begin{table}[htbp]
\centering
\begin{tabular}{Sl|Sc|Sc}
\hline
\textbf{Encoding} & \textbf{3 components} & \textbf{2 components} \\ \hline
Dynamical (Eq.~\eqref{eq:unitaryevolvedstate})
& $\dfrac{9}{N(N+2)t^2}$ & $\dfrac{4}{N(N+2)t^2}$ \\ \hline
Thermal (Eq.~\eqref{eq:Gibbsstate})
& $\dfrac{36}{N(N+2)\beta^2}$ & $\dfrac{16}{N(N+2)\beta^2}$ \\ \hline
GS, $\Delta>0$
& $\dfrac{9\Delta^2}{N(N+2)}$ & $\dfrac{4\Delta^2}{N(N+2)}$ \\ \hline
\end{tabular}
\caption{Lower bounds on $\mathrm{Tr}[WQ^{-1}]$ for the estimation of respectively three- and two-dimensional magnetic fields and in the presence of a cost matrix $W$, for different parameter encoding schemes. Note that since $\sum_{i=1}^d \sqrt{w_i}=d$, the bounds do not depend on the cost matrix, but the optimal configuration does.}
\label{tab:QFIM_fundamentalbounds_magnetometry}
\end{table}

We finally note that these bounds are not tight for $d=1$. Tight bounds for this case were derived in Refs.~\cite{Boixo_2007, Abiuso_2025} and scale as $N^{-2}$, whereas our bounds scale as $[N(N+2)]^{-1}$, with the same prefactors, and are therefore always looser.

\section{Optimal systems in vector magnetometry}
We now identify optimal systems for vector magnetometry (with probe-encoding Hamiltonian as in Eq.~\eqref{eq:hamiltonianmagnetometry}) in the case $W=\mathbb{I}_d$. This is the most natural choice, since the WMSE becomes the squared Euclidean distance between the estimated and the true parameter vector. 

In what follows we consider local estimation around $\theta_i \approx 0$. Otherwise, a known compensating field can bring the working point close to zero, leaving only a small unknown residual, so the optimal configurations extend to arbitrary field values. All configurations below satisfy the WCC and, with the only exception of two-parameter thermal metrology, saturate the bounds of Table~\ref{tab:QFIM_fundamentalbounds_magnetometry} exactly (for $N=8n$ in the three-parameter unitary and GS cases), prefactor included, i.e., they achieve $Q=\kappa\Gamma$ with a diagonal $\Gamma$ that exhausts the variance budget. For unitary and GS metrology the probes are pure, so
that the WCC guarantees local asymptotic attainability of the QCRB with projective
measurements~\cite{Pezze_2017}, given in Appendix~\ref{app:optimalsystems}; for the
thermal probes, which are mixed, optimal measurements are left for future work.

\subsection*{Optimal systems for unitary metrology}
Three-parameter bounds for this setting without a control Hamiltonian were established in Ref.~\cite{Hou_2020}: restricted to that case they are generally
tighter at finite fields, but controllability allows one to surpass them. Control-free
optimal probes were identified in Refs.~\cite{Hou_2020,Baumgratz_2016}; in particular,
for $H^C=0$ the optimal probe for the local estimation of a weak field is
\cite{Baumgratz_2016} $\ket{\psi} \coloneqq \mathcal{N}(\ket{\Phi_1}+\ket{\Phi_2}+\ket{\Phi_3})$, with $\mathcal{N}$ a normalization constant and
$\ket{\Phi_k}=(\ket{\phi^+_k}^{\otimes N}+\ket{\phi^-_k}^{\otimes N})/\sqrt{2}$ the
GHZ state along the $k$-th axis, $\ket{\phi^\pm_k}$ being the eigenvectors of
$\sigma_k$. We refer to $\ket{\psi}$ as the XYZ-GHZ state. For $N=8n$ its QFIM
saturates the three-parameter bound (see Appendix~\ref{app:optimalsystemthreeparunitary}), and otherwise departs from it only by terms
exponentially small in $N$; controllability extends the saturation to arbitrary
fields, and optimal measurements are given in Ref.~\cite{Baumgratz_2016}.

For two
components, say $\theta_x$ and $\theta_y$, the optimal configuration is $H^C=0$ with
the Dicke state $\ket{N/2,0}$, $N$ even (see Appendix~\ref{app:optimalsystemtwoparunitary});
since $\ket{N/2,0}$ is an eigenstate of $J_z$, the third row and column of the QFIM
vanish and ignorance of $\theta_z$ does not affect the other two.

\subsection*{Optimal systems for thermal equilibrium metrology}
The three-parameter bound is achieved in the limit
$M\gg\beta^{-1}$ by $H^C=-MJ^2$, with $M>0$ and $J^2=J_x^2+J_y^2+J_z^2$
(see Appendix~\ref{app:optimalsystemthreeparthermal}); the corresponding Gibbs state is the maximally mixed state on the symmetric (Dicke) subspace. Since $[H_{\vec{\theta}},J_i]=0$, Eq.~\eqref{eq:thermalsandwich} gives $Q=\beta^2\Gamma$, with $\Gamma=\frac{N(N+2)}{12}\mathbb{I}_3$.

For two components, consider $H^C=c\,J_z^2-MJ^2$ with $M\gg\beta^{-1}$. By symmetry, $\Gamma$ and $K$ are diagonal, with $\Gamma_{xx}=\Gamma_{yy}=(A_0-\braket{J_z^2}_\rho)/2$ and $\beta K_{xx}=\beta K_{yy}=\beta c\,(A_0-3\braket{J_z^2}_\rho)$, where $A_0=N(N+2)/4$ (see Appendix~\ref{app:transversed2}), and Eq.~\eqref{eq:thermalupper} gives
\begin{equation*}
  \mathrm{Tr}[Q^{-1}]\leq\frac{16}{\beta^2N(N+2)}\left[1-\frac{\beta c}{3}-(1-\beta c)\frac{\braket{J_z^2}_\rho}{A_0}\right]^{-1}.
\end{equation*}
A weaker coupling reduces the loss due to non-commutativity, $\beta c/3$, but increases $\braket{J_z^2}_\rho\simeq1/(2\beta c)$, the part of the variance budget spent on the direction that is not estimated. Balancing the two at $\beta c=\sqrt{6/[N(N+2)]}$ gives $\mathrm{Tr}[Q^{-1}]=\frac{16}{\beta^2N(N+2)}\big[1+4/\sqrt{6N(N+2)}+\mathcal{O}(N^{-2})\big]$, where the first-order correction is exact, as upper and lower bounds coincide to this order (see Appendix~\ref{app:scalingsd2}). The same asymptotic precision follows for any $\beta c\propto N^{-\gamma}$ with $0<\gamma<2$. Since $H^C$ is
even in $J_z$, the Gibbs state is invariant under a $\pi$ rotation about the
$x$ axis, under which $L_1$ is even and $L_2$ odd; $[L_1,L_2]$ is therefore
odd and the WCC is satisfied exactly for all $N$. Finally, the full
$3\times3$ QFIM is diagonal, and ignorance of $\theta_3$ does not affect the other two.

\subsection*{Optimal systems for GS metrology}
The three-parameter bound is achieved for $N=8n$ by $H^C= -(\Delta/2)\ket{\psi} \bra{\psi} +(\Delta /2)\sum_{i=1}^{3} \ket{\psi_{i}} \bra{\psi_{i}} + \sum_{i=4}^{\dim(\mathcal{H})-1} E_i\ket{\tilde{\psi}_{i}} \bra{\tilde{\psi}_{i}}$, where $\ket{\psi}$ is the XYZ-GHZ state, $\ket{\psi_{i}} =J_i \ket{\psi}/(\sqrt{\braket{\psi |J_i^2| \psi}})$ for all $i=1,2,3$, $E_i-E_0 \geq \Delta$ for all $i>3$ and $\{ \ket{\tilde{\psi}_i} \}_{i=4}^{\dim(\mathcal{H})-1}$ are chosen to be orthogonal to $\ket{\psi}$ and $\{\ket{\psi_{i}}\}$ (see Appendix~\ref{app:optimalsystemthreeparGS}).

For two components, $H^C=gJ_z^2-J^2$ with $N$ even and $0<g\leq N$ (so that $\Delta=g$) (already known to give quadratic QFI scaling in the
single-parameter case \cite{andre2026}) saturates the $d=2$ bound
(see Appendix~\ref{app:optimalsystemtwoparGS}); since it commutes with $J_z$, ignorance of
$\theta_z$ again does not affect the other two.
    
\section{Bounds on the rank of the QFIM}
Finally, we also notice that other figures of merit, beyond the WMSE (Eq.~\eqref{scalarQCRB}), can be meaningful to multiparameter metrology in different operational contexts \cite{Candeloro_2024,he2026fishergeometry, oconnor2026geometricobstructions}. In Appendix~\ref{app:RANK_MAX} we bound the rank of $Q$, which counts how many
independent parameters a $D$-dimensional probe can be sensitive to, and find a sharp separation: out of the $D^2-1$ traceless directions in which a Hamiltonian can be perturbed, thermal states can be sensitive to all, whereas ground states and unitarily evolved pure states are limited to $2(D-1)$. This makes thermal states well-suited for the simultaneous estimation of a large number of parameters, whereas ground states and unitarily evolved pure states are better suited to distributed private sensing~\cite{hassani2025privacy}, where a low rank of $Q$ is precisely the desired feature.

\section{Conclusions}
We have investigated the ultimate limits of multiparameter quantum metrology and derived general bounds on $\text{Tr}[WQ^{-1}]$ for three paradigmatic control-Hamiltonian encodings: unitary dynamics, thermal equilibrium, and ground states. These bounds generalize known single-parameter results \cite{Boixo_2007, Abiuso_2025}, while clarifying the stricter conditions required for their saturation in the multiparameter regime. Obtained by optimizing over the control Hamiltonian $H^C$, these bounds are model-independent and thus provide benchmarks for assessing many-body probes. This complements previous studies, which have mostly focused on specific models, interaction types, or critical scenarios \cite{Montenegro_2025, Mihailescu_2024, difresco2022multiparameter}. 

We have applied these results to vector magnetometry, deriving asymptotically tight bounds for the simultaneous estimation of two and three components of weak magnetic fields, and identifying unitary, ground-state, and thermal-state many-body probes that saturate them. In doing so, we extend earlier results in multiparameter unitary magnetometry \cite{Baumgratz_2016, Hou_2020, oconnor2026geometricobstructions}, and provide guidelines for the design of optimal many-body probes for vector field sensing.

Our work opens several avenues for future research. First, since our results rely on the local, asymptotic regime captured by the quantum Cramér--Rao bound, extending this framework to the finite-resource regime through Bayesian multiparameter strategies \cite{Kaubruegger2023, Bavaresco2024, albarelli2025measurement, Mihailescu2025, andre2026strategy} would be a natural step towards realistic sensing protocols. Second, it would be interesting to explore how these bounds are modified under adaptive or sequential estimation schemes \cite{Yang2025, Fazio2026}, where new trade-offs may emerge in thermal and ground-state metrology. Finally, a natural direction is to extend our approach to probes subject to experimental constraints or imperfections, where the interplay between control, noise, and measurement incompatibility remains largely unexplored \cite{roy2019fundamental_noisy_multiparameter_bounds, Albarelli_2022}.

\textit{Note added}---We note that, during the preparation of this manuscript, matrix bounds on the QFIM at thermal equilibrium, including the upper bound in Eq.~\eqref{eq:thermalsandwich}, were independently derived in Ref.~\cite{Cao_2026}, from which the same scalar bounds for general Hamiltonians can also be recovered.

\section*{Acknowledgments}
R.P. acknowledges the support of the Catalonia Quantum Academy through its Student Mobility Fellowships program. V. I. and M.P.-L. acknowledge funding from  the ATRAE Program (Grant ATR2024-154621) funded by MICIU/AEI/10.13039/501100011033.
 J.C. acknowledges
support from ICREA Academia. Additional support was provided by grant PID2022-141283NB-I00 (MICIU/AEI/10.13039/501100011033).
A.B. acknowledges the support of the Harish-Chandra Research Institute, Prayagraj through its Visiting Students Programme. P.A. acknowledges funding from the Austrian Science Fund (FWF) projects 10.55776/I6004 and 10.55776/ESP2889224.

\bibliography{bibliography_arXiv}

\onecolumngrid
\appendix

\section{QFIM for parametrized Gibbs states} \label{app:QFIMthermal}
Here we derive the expression for the QFIM at thermal equilibrium used in the main text.\\
Using the definition of the SLD, $\partial_i\rho_{\vec{\theta}}=\{L_i,\rho_{\vec{\theta}}\}/2=\mathbb{J}_{B,\rho}[L_i]$, the QFIM can be written as
$Q_{ij}=\mathrm{Tr}[\rho_{\vec{\theta}}\{L_i,L_j\}]/2=\mathrm{Tr}\big[\partial_i\rho_{\vec{\theta}}\,\mathbb{J}_{B,\rho}^{-1}[\partial_j\rho_{\vec{\theta}}]\big]$.
As shown in~\cite{Abiuso_2025}, the derivative of a Gibbs state is
$\partial_i\rho_{\vec{\theta}}=-\beta\mathbb{J}_{L,\rho}[\partial_iH_{\vec{\theta}}]+\beta\rho_{\vec{\theta}}\mathrm{Tr}[\rho_{\vec{\theta}}\partial_iH_{\vec{\theta}}]=-\beta\mathbb{J}_{L,\rho}[\tilde H_i]$,
with $\tilde H_i=\partial_iH_{\vec{\theta}}-\mathrm{Tr}[\rho_{\vec{\theta}}\partial_iH_{\vec{\theta}}]$, where we used $\mathbb{J}_{L,\rho}[\mathbb{I}]=\rho_{\vec{\theta}}$.
Since $\mathbb{J}_{L,\rho}$ is self-adjoint with respect to the Hilbert--Schmidt inner product~\cite{Abiuso_2025},
$Q_{ij}=\beta^2\mathrm{Tr}\big[\tilde H_i\,\mathcal{J}_\rho[\tilde H_j]\big]$, with
$\mathcal{J}_\rho = \mathbb{J}_{L,\rho} \circ \mathbb{J}^{-1}_{B,\rho} \circ \mathbb{J}_{L,\rho}$. Finally, $\mathcal{J}_\rho[\mathbb{I}]=\rho_{\vec{\theta}}$
and, by self-adjointness, $\mathrm{Tr}\big[\mathcal{J}_\rho[A]\big]=\mathrm{Tr}[\rho_{\vec{\theta}}A]$, which give
\begin{equation}
    Q_{ij} = \beta^2 \left( \mathrm{Tr}[\, \partial_i H_{\vec{\theta}} \, \mathcal{J}_{\rho}[\partial_j H_{\vec{\theta}}] \,] - \mathrm{Tr}[\rho_{\vec{\theta}} \partial_i H_{\vec{\theta}}]\,\mathrm{Tr}[\rho_{\vec{\theta}} \partial_j H_{\vec{\theta}}]\right).
    \label{eqS:QFIMthermal}
\end{equation}

\section{QFIM for parametrized non-degenerate ground states} \label{app:QFIMGS}
Starting from Eq.~\eqref{eqS:QFIMthermal}, we can express it in the low-temperature limit, thus generalizing to the multiparameter case the already known formula for the single-parameter QFI associated with parametrized non-degenerate ground states with gap $\Delta$ \cite{Abiuso_2025}:
\begin{equation}
    \mathcal{F}_\theta = \sum_{k>0} \frac{4 |(\partial_\theta H_\theta)_{0k}|^2}{(E_k - E_0)^2} + \mathcal{O}(\beta^2e^{-\beta\Delta}),
    \label{eq:QFIlowT}
\end{equation}
where $E_k$ denotes the $k$-th eigenvalue of $H_\theta$, ordered in increasing order, and $(\cdot)_{jk}$ denotes matrix elements in the eigenbasis of $\rho_\theta$, or equivalently of $H_\theta$.

To this purpose, we start by observing that, given an operator $A$ \cite{Abiuso_2025},
\begin{equation*}
    \mathcal{J}_\rho[A]
    =
    \sum_{\substack{i,j \\ p_i \neq p_j}}
    \frac{2(p_i-p_j)^2}{(\ln p_i - \ln p_j)^2 (p_i + p_j)} A_{ij} \ket{i}\bra{j}
    +
    \sum_{\substack{i,j \\ p_i=p_j}} p_i A_{ij} \ket{i}\bra{j}.
\end{equation*}

We now rewrite $\rho_{\vec{\theta}}$ as
\begin{equation*}
    \rho_{\vec{\theta}} = \sum_i p_i \ket{i}\bra{i} = \sum_a a\, \Pi_a,
\end{equation*}
where $\Pi_a = \sum_{i\,|\,p_i=a} \ket{i}\bra{i}$ are the projectors onto the eigenspaces of $\rho_{\vec{\theta}}$ with eigenvalue $a$. It is straightforward to see that
\begin{align*}
    \mathcal{J}_\rho [\Pi_a A \Pi_a] &= a\, \Pi_a A \Pi_a, \\
    \mathcal{J}_\rho [\Pi_a A \Pi_b] &= \frac{2(a-b)^2}{(\ln a - \ln b)^2 (a+b)} \Pi_a A \Pi_b.
\end{align*}

We now observe that
\begin{align*}
    \mathrm{Tr}[\partial_i H_{\vec{\theta}} \mathcal{J}_\rho[\partial_j H_{\vec{\theta}}]]
    &=
    \sum_{a,b}
    \mathrm{Tr}[\partial_i H_{\vec{\theta}} \mathcal{J}_\rho[\Pi_b \partial_j H_{\vec{\theta}} \Pi_a]] \\
    &=
    \mathrm{Tr}\!\left[\sum_a a\, \partial_i H_{\vec{\theta}} \Pi_a \partial_j H_{\vec{\theta}} \Pi_a \right]
    +
    \mathrm{Tr}\!\left[\sum_{a \neq b} \frac{2(a-b)^2}{(\ln a - \ln b)^2 (a+b)} \partial_i H_{\vec{\theta}} \Pi_b \partial_j H_{\vec{\theta}} \Pi_a \right], \\
    \mathrm{Tr}[\rho_{\vec{\theta}} \partial_i H_{\vec{\theta}}]
    &=
    \mathrm{Tr}\!\left[\sum_a a\, \Pi_a \partial_i H_{\vec{\theta}} \right].
\end{align*}

Substituting these expressions into Eq.~\eqref{eqS:QFIMthermal}, we find
\begin{align*}
    Q_{ij}\beta^{-2}
    &=
    \mathrm{Tr}\!\left[\sum_a a\, \partial_i H_{\vec{\theta}} \Pi_a \partial_j H_{\vec{\theta}} \Pi_a \right]
    -
    \mathrm{Tr}\!\left[\sum_a a\, \Pi_a \partial_i H_{\vec{\theta}} \right]
    \mathrm{Tr}\!\left[\sum_b b\, \Pi_b \partial_j H_{\vec{\theta}} \right] \\
    &\quad +
    \mathrm{Tr}\!\left[\sum_{a \neq b} \frac{2(a-b)^2}{(\ln a - \ln b)^2 (a+b)} \partial_i H_{\vec{\theta}} \Pi_b \partial_j H_{\vec{\theta}} \Pi_a \right].
\end{align*}

We now perform the low-temperature limit, where only the ground state is populated, with probability $p_0 = 1$. The first term reads
\begin{align*}
    \mathrm{Tr}\!\left[\sum_a a \sum_{l\,|\,p_l=a} \ket{l}\bra{l}\partial_i H_{\vec{\theta}} \sum_{m\,|\,p_m=a} \ket{m}\bra{m}\partial_j H_{\vec{\theta}} \right]
    & =
    \sum_a a \sum_{l\,|\,p_l=a} \sum_{m\,|\,p_m=a}
    \bra{l}\partial_i H_{\vec{\theta}} \ket{m}
    \bra{m}\partial_j H_{\vec{\theta}} \ket{l} \\
    &\qquad =
    \bra{0}\partial_i H_{\vec{\theta}} \ket{0}
    \bra{0}\partial_j H_{\vec{\theta}} \ket{0}
    + \mathcal{O}(e^{-\beta \Delta}).
\end{align*}

Similarly, the second term reads
\begin{align*}
    -\mathrm{Tr}\!\left[\sum_a a \sum_{l\,|\,p_l=a} \ket{l}\bra{l}\partial_i H_{\vec{\theta}} \right]
    \mathrm{Tr}\!\left[\sum_b b \sum_{m\,|\,p_m=b} \ket{m}\bra{m} \partial_j H_{\vec{\theta}} \right] = -
    \bra{0}\partial_i H_{\vec{\theta}} \ket{0}
    \bra{0}\partial_j H_{\vec{\theta}} \ket{0}
    + \mathcal{O}(e^{-\beta \Delta}).
\end{align*}

Thus, in the low-temperature limit, the first two terms cancel each other. We now analyze the last term:
\begin{align*}
    \mathrm{Tr}\!\left[\sum_{a \neq b} \frac{2(a-b)^2}{(\ln a - \ln b)^2 (a+b)} \partial_i H_{\vec{\theta}} \Pi_b \partial_j H_{\vec{\theta}} \Pi_a \right] &=
    \sum_{a \neq b} \frac{2(a-b)^2}{(\ln a - \ln b)^2 (a+b)}
    \sum_{l\,|\,p_l=a} \sum_{m\,|\,p_m=b}
    \bra{l}\partial_i H_{\vec{\theta}} \ket{m}
    \bra{m}\partial_j H_{\vec{\theta}} \ket{l} \\
    &=
    \sum_{b>0} \frac{2}{\beta^2 (E_b - E_0)^2}
    \sum_{k\,|\,p_k=b}
    \bra{0}\partial_i H_{\vec{\theta}} \ket{k}
    \bra{k}\partial_j H_{\vec{\theta}} \ket{0} \\
    &\quad +
    \sum_{a>0} \frac{2}{\beta^2 (E_a - E_0)^2}
    \sum_{k\,|\,p_k=a}
    \bra{k}\partial_i H_{\vec{\theta}} \ket{0}
    \bra{0}\partial_j H_{\vec{\theta}} \ket{k} \\
    &=
    \sum_{b>0} \frac{4}{\beta^2 (E_b - E_0)^2}
    \sum_{k\,|\,p_k=b}
    \mathrm{Re}\!\left[
    \bra{0}\partial_i H_{\vec{\theta}} \ket{k}
    \bra{k}\partial_j H_{\vec{\theta}} \ket{0}
    \right],
\end{align*}
where we used $\ln p_k-\ln p_0=-\beta(E_k-E_0)$. We thus obtain
\begin{equation}
    Q_{ij} = \sum_{k>0} \frac{4\,\mathrm{Re}\!\left[\bra{0}\partial_i H_{\vec{\theta}} \ket{k}\bra{k}\partial_j H_{\vec{\theta}}\ket{0}\right]}{(E_k - E_0)^2} + \mathcal{O}(\beta^2e^{-\beta\Delta}),
    \label{eqS:QFIMGS}
\end{equation}
where we have expressed $H_{\vec{\theta}} = \sum_k E_k \ket{k}\bra{k}$.

\section{General bounds in multiparameter quantum metrology}
In this section we derive general bounds for multiparameter quantum metrology with unitary dynamics, at
thermal equilibrium, and for ground states. They all follow from a common matrix structure: the QFIM is
upper bounded, in the sense of positive semidefinite matrices, by a covariance matrix of generators,
\begin{equation}
  Q\le\kappa\,\Gamma,\qquad\Gamma_{ij}=\mathrm{Cov}_\rho(G_i,G_j),
  \label{eqS:QleqGamma}
\end{equation}
with $G_i=\mathcal{H}_i$ and $\kappa=4$ for unitary dynamics (Sec.~\ref{app:boundunitary}),
$G_i=\partial_iH_{\vec{\theta}}$ and $\kappa=\beta^2$ at thermal equilibrium (Sec.~\ref{app:boundthermal}),
and $G_i=\partial_iH_{\vec{\theta}}$ and $\kappa=4/\Delta^2$ for ground states (Sec.~\ref{app:boundGS}), the
covariances being evaluated in the initial state, the Gibbs state, and the ground state, respectively.
Since matrix inversion reverses the order between positive definite matrices, Eq.~\eqref{eqS:QleqGamma}
implies $Q^{-1}\ge\kappa^{-1}\Gamma^{-1}$. Moreover, any positive definite matrix $A$ satisfies
$(A^{-1})_{ii}\ge1/A_{ii}$~\cite{Albarelli_2022}, so that, for diagonal $W$,
\begin{equation}
    \mathrm{Tr}[WA^{-1}] \geq \sum_{i=1}^d \frac{w_i}{A_{ii}},
    \label{eqS:boundtraceQFIM}
\end{equation}
with equality whenever $A$ is diagonal. Applying it to $A=\Gamma$, we obtain
\begin{equation}
  \mathrm{Tr}[WQ^{-1}]\ge\frac{1}{\kappa}\mathrm{Tr}[W\Gamma^{-1}]\ge\frac{1}{\kappa}\sum_{i=1}^d\frac{w_i}{\Gamma_{ii}}.
  \label{eqS:matrixchain}
\end{equation}
The first inequality is saturated if and only if $Q=\kappa\Gamma$, and the second one whenever $\Gamma$ is
diagonal. In each of the following subsections we prove Eq.~\eqref{eqS:QleqGamma} and combine it with
Eq.~\eqref{eqS:matrixchain}; probe-independent bounds then follow by maximizing the variances $\Gamma_{ii}$
(Sec.~\ref{app:upperboundvariance}). At thermal equilibrium, Eq.~\eqref{eqS:QleqGamma} is complemented by a
lower bound (Sec.~\ref{app:thermalsandwich}), which yields an upper bound on $\mathrm{Tr}[WQ^{-1}]$ for a
given probe.

\subsection{Upper bound on the variance of a Hermitian operator and saturability in multiparameter quantum metrology} \label{app:upperboundvariance}
For any state $\rho$ and Hermitian operator $A$~\cite{Abiuso_2025},
\begin{equation}
\mathrm{Var}_\rho(A)\le \frac{\|A\|^2}{4},
\label{eq:boundvariance}
\end{equation}
since the variance of a probability distribution supported on $[\lambda_\mathrm{min}(A),\lambda_\mathrm{max}(A)]$
is at most $\|A\|^2/4$. Equality holds if and only if the outcome distribution of $A$ in $\rho$ assigns
probability $1/2$ to each of the extremal eigenvalues, as for the pure state
$(\ket{\lambda_\mathrm{max}}+e^{i\phi}\ket{\lambda_\mathrm{min}})/\sqrt2$ or the mixed state
$(\ket{\lambda_\mathrm{max}}\bra{\lambda_\mathrm{max}}+\ket{\lambda_\mathrm{min}}\bra{\lambda_\mathrm{min}})/2$,
where $\ket{\lambda_\mathrm{max}}$ ($\ket{\lambda_\mathrm{min}}$) is an eigenvector of $A$ with maximal (minimal)
eigenvalue.

The probe-independent bounds in Eqs.~\eqref{eqS:boundunitary}, \eqref{eqS:boundthermal}
and~\eqref{eq:inequalitiesGSmagnetometry} are therefore saturated only if a single state satisfies this
condition for all generators at once: the initial state for the operators $\mathcal{H}_i$ in unitary
metrology, and the Gibbs state (which can only approach the mixed form above) or the ground state for
the derivatives $\partial_iH_{\vec{\theta}}$. Such a state need not exist, even for commuting generators.
For example, for $H_{\vec{\theta}}=\theta_1J_z+\theta_2J^2$ with $N$ even, the extremal eigenspaces of $J_z$
belong to the sector $j=N/2$, whereas the minimal eigenspace of $J^2$ is the sector $j=0$, so that no state
maximizes both variances. Conversely, for the non-commuting generators $\sigma_x$ and $\sigma_y$, the state
$\ket{0}=(\ket{+}+\ket{-})/\sqrt{2}=(\ket{+y}+\ket{-y})/\sqrt{2}$ maximizes both.

\subsection{General bounds in multiparameter unitary metrology} \label{app:boundunitary}
Let us now focus on unitary quantum metrology. In this setting, parameter estimation can be enhanced by both optimizing over the initial quantum state $\rho_0$ and the control Hamiltonian $H^C$. We first keep the control Hamiltonian fixed and optimize over the initial state. Given a state parametrized as $\rho_{\vec{\theta}} = U_{\vec{\theta}} \rho_0 U_{\vec{\theta}}^\dagger$, with $U_{\vec{\theta}} = e^{-i t H_{\vec{\theta}}/\hbar}$, the QFIM entries are given by \cite{Liu_2019}:
\begin{align}
Q_{ij} =
\sum_{p_k \neq 0} 4p_k \, 
\mathrm{Cov}_{\ket{p_k}}(\mathcal{H}_i,\mathcal{H}_j) 
- \sum_{\substack{p_k,p_l \neq 0 \\ k \neq l}}
\frac{8 p_k p_l}{p_k + p_l}
\mathrm{Re}\big(
\braket{p_k|\mathcal{H}_i|p_l}
\braket{p_l|\mathcal{H}_j|p_k}
\big), \label{eq:QFIMunitary}
\end{align}
where $\rho_0=\sum_k p_k \ket{p_k}\bra{p_k}$, $ \mathcal{H}_i \coloneqq i (\partial_i U_{\vec{\theta}}^\dagger)U_{\vec{\theta}}$, and the covariance of two operators $A,B$ under a quantum state $\rho$ is defined as
\begin{equation*}
    \mathrm{Cov}_\rho({A,B})=\frac{1}{2}\mathrm{Tr}_\rho\left[\{A,B\}\right] -\mathrm{Tr}_\rho\left[A\right]\mathrm{Tr}_\rho\left[B\right].
\end{equation*}
The second term in Eq.~\eqref{eq:QFIMunitary} is a positive semidefinite matrix: for any real vector $\vec u$
and $\mathcal{H}_{\vec u}=\sum_iu_i\mathcal{H}_i$, its quadratic form is
$\sum_{k\neq l}\frac{8p_kp_l}{p_k+p_l}|\braket{p_k|\mathcal{H}_{\vec u}|p_l}|^2\ge0$. Hence
$Q\le4\sum_kp_k\,\mathrm{Cov}_{\ket{p_k}}(\mathcal{H})$, where $\mathrm{Cov}_\rho(\mathcal{H})$ denotes the
covariance matrix with elements $\mathrm{Cov}_\rho(\mathcal{H}_i,\mathcal{H}_j)$. Moreover, the covariance
matrix is concave in the state,
$\mathrm{Cov}_{\rho_0}(\mathcal{H})-\sum_kp_k\mathrm{Cov}_{\ket{p_k}}(\mathcal{H})=\sum_kp_k(\vec m_k-\vec m)(\vec m_k-\vec m)^T\ge0$,
with $(\vec m_k)_i=\braket{p_k|\mathcal{H}_i|p_k}$ and $\vec m=\sum_kp_k\vec m_k$. Therefore
$Q\le4\,\mathrm{Cov}_{\rho_0}(\mathcal{H})$, which is Eq.~\eqref{eqS:QleqGamma} with $\kappa=4$, with equality
for pure initial states. Combining it with Eqs.~\eqref{eqS:matrixchain} and~\eqref{eq:boundvariance},
\begin{align}
    \mathrm{Tr}[WQ^{-1}] \geq \frac{1}{4}\mathrm{Tr}[W\Gamma^{-1}]
    \geq \min_{\rho_0}\sum_{i=1}^d \frac{w_i}{4\mathrm{Var}_{\rho_0} (\mathcal{H}_i)}\geq\sum_{i=1}^d \frac{w_i}{\| \mathcal{H}_i \|^2}.
\end{align}
The first inequality is saturated for pure initial states, the second one requires the optimal initial
state to have a diagonal $\Gamma$, and the last one a state that maximizes all the variances
simultaneously (Sec.~\ref{app:upperboundvariance}).

We now optimize over the control Hamiltonian. We start by observing that $\mathcal{H}_i$ can be expressed as \cite{Liu_2019}
\[
    \mathcal{H}_i \coloneqq i (\partial_i U_{\vec{\theta}}^\dagger)U_{\vec{\theta}}
    =
    -\int_0^1 ds \,
    e^{isH_{\vec{\theta}} t/\hbar}
    \frac{t}{\hbar}\partial_i H_{\vec{\theta}}
    e^{-isH_{\vec{\theta}} t/\hbar}
    = - \int_0^1 ds \, \frac{t}{\hbar} H'^{{(s)}}_i ,
\]
where we define the partially rotated $i$-th derivative Hamiltonian as
\begin{equation*}
    H'^{{(s)}}_i \coloneqq
    e^{isH_{\vec{\theta}} t/\hbar}
    \partial_i H_{\vec{\theta}}
    e^{-isH_{\vec{\theta}} t/\hbar}.
\end{equation*}
Since the variance is convex in its operator argument
\begin{equation}
    \mathrm{Var}_\rho(\mathcal{H}_i) \le  \int_0^1 ds \, \frac{t^2}{\hbar^2}\mathrm{Var}_\rho(H'^{{(s)}}_i). \label{eq:boundvariancerotatedderivative}
\end{equation}
Therefore
\begin{equation*}
    \max_\rho \mathrm{Var}_\rho(\mathcal{H}_i) \leq \frac{t^2}{\hbar^2} \int_0^1ds\, \max_\rho \mathrm{Var}_\rho(H'^{{(s)}}_i) \le \frac{t^2}{\hbar^2} \max_s\max_\rho \mathrm{Var}_\rho(H'^{{(s)}}_i).
\end{equation*}
By unitary invariance
\[
     \max_\rho \mathrm{Var}_\rho(H'^{{(s)}}_i) = \max_\rho \mathrm{Var}_\rho(\partial_i H_{\vec{\theta}}).
\]
Therefore
\[
        \frac{\| \mathcal{H}_i \|^2}{4}=\max_\rho \mathrm{Var}_\rho(\mathcal{H}_i) \le \frac{t^2}{\hbar^2}\max_\rho\mathrm{Var}_\rho(\partial_i H_{\vec{\theta}}) =\frac{t^2 \| \partial_i H_{\vec{\theta}}\|^2}{4 \hbar^2}.
    \]
Combining this with the previous chain, we finally obtain:
\begin{align}
    \mathrm{Tr}[WQ^{-1}] \geq  \frac{1}{4}\mathrm{Tr}[W\Gamma^{-1}]
    \geq \min_{\rho_0}\sum_{i=1}^d \frac{w_i}{4\mathrm{Var}_{\rho_0} (\mathcal{H}_i)}\geq\sum_{i=1}^d \frac{w_i}{\| \mathcal{H}_i \|^2}\geq \sum_{i=1}^d \frac{w_i\hbar^2}{t^2 \|\partial_i H_{\vec{\theta}}\|^2}. \label{eqS:boundunitary}
\end{align}
The last inequality is saturated whenever $[\partial_i H_{\vec{\theta}}, H_{\vec{\theta}}]=0$.

\subsection{Convexity of the variance in its operator argument}
For completeness, here we prove Eq.~\eqref{eq:boundvariancerotatedderivative}. The proof relies on the well-known inequality
\begin{equation*}
    |\mathrm{Cov}_\rho(A,B)| \le \sqrt{\mathrm{Var}_\rho(A)\mathrm{Var}_\rho(B)}
    \quad \text{for all Hermitian operators } A,B.
\end{equation*}

From this, for any family of Hermitian operators \(A(s)\), it follows that
\begin{align*}
    \mathrm{Var}_\rho\left( \int_0^1 ds\, A(s) \right)
    &= \mathrm{Tr}\!\left[ \rho \int_0^1 ds\int_0^1dt\, A(s)A(t)\right]
    - \mathrm{Tr}\!\left[ \rho\int_0^1ds\,A(s) \right]^2 \\
    &= \int_0^1 ds\int_0^1dt\, \mathrm{Cov}_\rho(A(s),A(t))\\
    &\le \int_0^1 ds\int_0^1dt\, \sqrt{\mathrm{Var}_\rho(A(s))\mathrm{Var}_\rho(A(t))}\\
    &= \left(\int_0^1 ds\, \sqrt{\mathrm{Var}_\rho(A(s))}\right)^2.
\end{align*}

We now apply the Cauchy–Schwarz inequality for integrals,
\begin{equation*}
    \left|\int dx\, f(x) g^*(x)\right|^2 \le \int dx\, |f(x)|^2 \int dx\, |g(x)|^2.
\end{equation*}

Setting \(g(x)=1\), we obtain
\[
    \left(\int_0^1 ds\, \sqrt{\mathrm{Var}_\rho(A(s))}\right)^2
    \le
    \int_0^1 ds\, \mathrm{Var}_\rho(A(s)).
\]

Therefore,
\[
    \mathrm{Var}_\rho\left( \int_0^1 ds\, A(s) \right)
    \le
    \int_0^1 ds\, \mathrm{Var}_\rho(A(s)),
\]
which immediately implies Eq.~\eqref{eq:boundvariancerotatedderivative}.

\vspace{2mm}

More generally, the variance is convex in its operator argument:
\begin{align*}
    \mathrm{Var}_\rho(pA + (1-p)B)
    &= \langle (pA + (1-p)B)^2\rangle_\rho
    - \langle (pA + (1-p)B)\rangle_\rho^2\\
    &= p^2\mathrm{Var}_\rho(A)
    + (1-p)^2\mathrm{Var}_\rho(B)
    + 2p(1-p)\mathrm{Cov}_\rho(A,B)\\
    &\le p^2\mathrm{Var}_\rho(A)
    + (1-p)^2\mathrm{Var}_\rho(B)
    + 2p(1-p)\sqrt{\mathrm{Var}_\rho(A)\mathrm{Var}_\rho(B)}\\
    &= \left(p\sqrt{\mathrm{Var}_\rho(A)} + (1-p)\sqrt{\mathrm{Var}_\rho(B)}\right)^2\\
    &\le p\,\mathrm{Var}_\rho(A) + (1-p)\,\mathrm{Var}_\rho(B),
\end{align*}
where in the last step we used the convexity of the function \(x^2\).

\subsection{General bounds in multiparameter thermal equilibrium metrology}\label{app:boundthermal}
Let us now consider thermal equilibrium metrology. For any real vector $\vec u$, let
$V_{\vec u}=\sum_iu_i\partial_iH_{\vec{\theta}}$ and $\tilde V_{\vec u}=V_{\vec u}-\braket{V_{\vec u}}_\rho$. By
Eq.~\eqref{eqS:QFIMthermal} and the superoperator inequality
$\mathcal{J}_{\rho}\leq\mathbb{J}_{B,\rho}$~\cite{Abiuso_2025,scandi2024quantumfisherinformationdynamical,PETZ_2011},
\begin{equation*}
  \vec u^{\,T}Q\,\vec u=\beta^2\,\mathrm{Tr}\big[\tilde V_{\vec u}\,\mathcal{J}_\rho[\tilde V_{\vec u}]\big]
  \le\beta^2\,\mathrm{Tr}\big[\tilde V_{\vec u}\,\mathbb{J}_{B,\rho}[\tilde V_{\vec u}]\big]=\beta^2\,\mathrm{Var}_\rho(V_{\vec u}),
\end{equation*}
i.e., $Q\le\beta^2\Gamma$ with $\Gamma_{ij}=\mathrm{Cov}_\rho(\partial_iH_{\vec{\theta}},\partial_jH_{\vec{\theta}})$, which is
Eq.~\eqref{eqS:QleqGamma} with $\kappa=\beta^2$; its diagonal elements were obtained
in~\cite{GarciaPintos2024,Abiuso_2025} and the matrix form in~\cite{Cao_2026}. Combining it with
Eqs.~\eqref{eqS:matrixchain} and~\eqref{eq:boundvariance},
\begin{align}
    \mathrm{Tr}[WQ^{-1}] \geq \frac{1}{\beta^2}\mathrm{Tr}[W\Gamma^{-1}]
    \geq \sum_{i=1}^d \frac{w_i}{\beta^2 \mathrm{Var}_{\rho}(\partial_i H_{\vec{\theta}})}
    \geq \min_{H^C }\sum_{i=1}^d \frac{w_i}{\beta^2 \mathrm{Var}_{\rho}(\partial_i H_{\vec{\theta}})}
    \geq \sum_{i=1}^d \frac{4w_i}{\beta^2 \|\partial_i H_{\vec{\theta}}\|^2}. \label{eqS:boundthermal}
\end{align}

\subsubsection{\texorpdfstring{A two-sided bound on the thermal QFIM}{A two-sided bound on the thermal QFIM}}\label{app:thermalsandwich}

The first inequality in Eq.~\eqref{eqS:boundthermal}, which follows from
$Q\le\beta^2\Gamma$, can be complemented by a lower bound that
quantifies how far it is from saturation. The logic is as follows: a two-sided inequality
for a scalar kernel (Lemma~1) yields a two-sided matrix bound on the thermal QFIM (Theorem~1), from
which we obtain criteria for exact and asymptotic saturation (Corollaries~1 and~2); these are applied
to two-component thermal magnetometry in Sec.~\ref{app:optimalsystemtwoparthermal}. In this subsection we write
$H\equiv H_{\vec{\theta}}$ and $\rho\equiv\rho_{\vec{\theta}}$, and, for a single
parameter entering as $H+\theta V$ with $V$ Hermitian, we denote by $Q(V)$ the
corresponding thermal QFI, given by Eq.~\eqref{eqS:QFIMthermal}. We first record two
facts about the logarithmic mean.

\paragraph*{Lemma 1.}
For $p,q>0$ let $\ell=\ln(p/q)$, $A=\frac{p+q}{2}$ and $\Lambda=\frac{p-q}{\ell}$
(the logarithmic mean; $\Lambda=p$ if $p=q$). Then
\begin{equation}
  A-\frac{\Lambda^2}{A}=(p-q)\,g(\ell),
  \qquad
  g(\ell)=\frac12\coth\frac{\ell}{2}-\frac{2}{\ell^2}\tanh\frac{\ell}{2}
  =\frac{\ell}{6}-\frac{7\ell^3}{720}+\frac{53\ell^5}{60480}-\cdots ,
  \label{eqS:gexact}
\end{equation}
and, for all $\ell$,
\begin{equation}
  \frac{(p-q)\,\ell}{6}-\frac{(p-q)\,\ell^3}{96}\;\le\;A-\frac{\Lambda^2}{A}\;\le\;\frac{(p-q)\,\ell}{6}.
  \label{eqS:kernel}
\end{equation}
Moreover, $A-\Lambda^2/A\ge0$.

\emph{Proof.} The exact form follows by writing $p=re^{\ell/2}$ and $q=re^{-\ell/2}$, so
that $p-q=2r\sinh\frac{\ell}{2}$, $A=r\cosh\frac{\ell}{2}$ and $\Lambda=\frac{2r}{\ell}\sinh\frac{\ell}{2}$.
For the bounds, let $f(s)=p^s q^{1-s}$, so that $\Lambda=\int_0^1f(s)\,ds$,
$A=[f(0)+f(1)]/2$ and $f''=\ell^2f$. Integrating by parts twice against the
trapezoid kernel $s(1-s)/2$, and then once more using
$s(1-s)-\frac16=\frac{d^2}{ds^2}\big[-s^2(1-s)^2/12\big]$
(all boundary terms vanish), one obtains the exact identity
\begin{equation}
  A-\Lambda=\frac{\ell^2}{2}\int_0^1s(1-s)f\,ds
  =\frac{\ell^2}{12}\Lambda-\frac{\ell^4}{24}\int_0^1s^2(1-s)^2f\,ds .
  \label{eqS:ident}
\end{equation}
Since $s(1-s)\le\frac14$, the last integral lies between $0$ and
$\frac14\int_0^1s(1-s)f\,ds$, and Eq.~\eqref{eqS:ident} itself gives
$\int_0^1s(1-s)f\,ds=2(A-\Lambda)/\ell^2\le\Lambda/6$. Hence $a\equiv A-\Lambda$ satisfies
\begin{equation}
  \frac{\ell^2\Lambda}{12}\Big(1-\frac{\ell^2}{48}\Big)\;\le\;a\;\le\;\frac{\ell^2\Lambda}{12}.
  \label{eqS:abounds}
\end{equation}
Finally $A-\Lambda^2/A=(A-\Lambda)(A+\Lambda)/A=2a-a^2/A$, with
$0\le a^2/A\le a^2/\Lambda\le\ell^4\Lambda/144$. Inserting Eq.~\eqref{eqS:abounds} and
$\Lambda\ell=p-q$ gives the two bounds in Eq.~\eqref{eqS:kernel}, while
$A-\Lambda^2/A\ge0$ is $\Lambda\le A$, i.e., $a\ge0$ in the first form of Eq.~\eqref{eqS:ident}.
Note that Eq.~\eqref{eqS:kernel} reads $\frac{\ell}{6}-\frac{\ell^3}{96}\le g(\ell)\le\frac{\ell}{6}$
for $\ell\ge0$ ($g$ is odd); the lower constant is close to the best possible one,
$\frac{7}{720}$, which is approached as $\ell\to0$, and the bounds are proved without using the series. \hfill$\square$

\paragraph*{Theorem 1.}
Let $\mathrm{ad}_H(V)=[H,V]$. In an eigenbasis $\{E_k,\ket{k}\}$ of $H$, $(\mathrm{ad}_HV)_{kl}=(E_k-E_l)V_{kl}$,
and, for any function $\phi$, $\phi(\beta\,\mathrm{ad}_H)V$ denotes the operator with matrix elements
$\phi(\beta(E_k-E_l))V_{kl}$. With $g$ as in Eq.~\eqref{eqS:gexact},
\begin{equation}
  Q(V)=\beta^2\mathrm{Var}_\rho(V)-\beta^2\braket{\big[V,\,g(\beta\,\mathrm{ad}_H)V\big]}_\rho
  \qquad\text{(exactly)}.
  \label{eqS:exactQ}
\end{equation}
Consequently, with $K_n(V)=\braket{[V,\mathrm{ad}_H^n(V)]}_\rho\ge0$ for odd $n$,
\begin{equation}
  \beta^2\mathrm{Var}_\rho(V)-\frac{\beta^3}{6}K_1(V)\;\le\;Q(V)\;\le\;\beta^2\mathrm{Var}_\rho(V),
  \label{eqS:sandwichgen}
\end{equation}
with equality on the right if and only if $[H,V]=0$, and more precisely
\begin{equation}
  Q(V)=\beta^2\mathrm{Var}_\rho(V)-\frac{\beta^3}{6}K_1(V)+R_V,\qquad
  0\le R_V\le\frac{\beta^5}{96}K_3(V) .
  \label{eqS:lemmaQ}
\end{equation}
For the full QFIM, let $\Gamma_{ij}=\mathrm{Cov}_\rho(\partial_iH,\partial_jH)$ and
$(K_n)_{ij}=\braket{[\partial_iH,\mathrm{ad}_H^n(\partial_jH)]}_\rho$. Then, in the sense of
positive semidefinite matrices,
\begin{equation}
  \beta^2\Gamma-\frac{\beta^3}{6}K_1\;\le\;Q\;\le\;\beta^2\Gamma,
  \qquad
  Q\;\le\;\beta^2\Gamma-\frac{\beta^3}{6}K_1+\frac{\beta^5}{96}K_3 ,
  \label{eqS:matrix}
\end{equation}
with $Q=\beta^2\Gamma$ if and only if $[H,\partial_iH]=0$ for all $i$.

\emph{Proof.} \emph{Step 1 (spectral form).} Let $\{E_k,\ket{k}\}$ be an eigenbasis
of $H$ and $p_k=e^{-\beta E_k}/Z$, and define
\begin{equation}
  A_{kl}=\frac{p_k+p_l}{2},\qquad \Lambda_{kl}=\frac{p_k-p_l}{\ln p_k-\ln p_l}.
  \label{eqS:means}
\end{equation}
In this basis $\mathbb{J}_{B,\rho}$ multiplies matrix elements by $A_{kl}$ and
$\mathcal{J}_\rho$ by $\Lambda_{kl}^2/A_{kl}$ (Sec.~\ref{app:QFIMGS}),
so that Eq.~\eqref{eqS:QFIMthermal} reads $Q(V)=\beta^2\sum_{kl}|V_{kl}|^2\Lambda_{kl}^2/A_{kl}-\beta^2\braket{V}_\rho^2$,
while $\mathrm{Var}_\rho(V)=\sum_{kl}|V_{kl}|^2A_{kl}-\braket{V}_\rho^2$. Therefore
\begin{equation}
  \beta^2\mathrm{Var}_\rho(V)-Q(V)=\beta^2\sum_{k,l}|V_{kl}|^2\left(A_{kl}-\frac{\Lambda_{kl}^2}{A_{kl}}\right).
  \label{eqS:spectral}
\end{equation}
\emph{Step 2 (sum rule for odd functions).} The superoperator $\mathrm{ad}_H$ is
diagonal on $\ket{k}\bra{l}$ with eigenvalue $E_k-E_l$, so
$(\phi(\beta\,\mathrm{ad}_H)V)_{kl}=\phi(\beta(E_k-E_l))V_{kl}$ for any function $\phi$. If $\phi$ is odd,
\begin{equation}
  \begin{aligned}
  \braket{[V,\phi(\beta\,\mathrm{ad}_H)V]}_\rho
  &=\sum_{k,l}p_k|V_{kl}|^2\Big[\phi\big(\beta(E_l-E_k)\big)-\phi\big(\beta(E_k-E_l)\big)\Big]
  =2\sum_{k,l}p_k|V_{kl}|^2\,\phi\big(\beta(E_l-E_k)\big)\\
  &=\sum_{k,l}|V_{kl}|^2\,(p_k-p_l)\,\phi\big(\beta(E_l-E_k)\big),
  \end{aligned}
  \label{eqS:sumrule}
\end{equation}
the last equality by relabeling $k\leftrightarrow l$ in half of the sum. For
$\phi(y)=y^n$ this gives $K_n(V)=\sum_{kl}|V_{kl}|^2(p_k-p_l)(E_l-E_k)^n$, whose terms are
all nonnegative because $p_k-p_l$ and $E_l-E_k$ have the same sign for a Gibbs
state; hence $K_n(V)\ge0$.
\emph{Step 3 (combination).} Apply Lemma~1 to each pair $(p_k,p_l)$, with
\begin{equation}
  \ell_{kl}=\ln\frac{p_k}{p_l}=\beta(E_l-E_k).
  \label{eqS:ellkl}
\end{equation}
The exact form~\eqref{eqS:gexact} turns Eq.~\eqref{eqS:spectral} into
$\beta^2\sum_{kl}|V_{kl}|^2(p_k-p_l)\,g(\beta(E_l-E_k))$, which is Eq.~\eqref{eqS:exactQ}
by Eq.~\eqref{eqS:sumrule} with $\phi=g$. The bounds~\eqref{eqS:kernel}, with
$(p_k-p_l)\ell_{kl}^n=\beta^n(p_k-p_l)(E_l-E_k)^n$, give
\begin{equation}
  \frac{\beta^3}{6}K_1(V)-\frac{\beta^5}{96}K_3(V)\;\le\;\beta^2\mathrm{Var}_\rho(V)-Q(V)\;\le\;\frac{\beta^3}{6}K_1(V) ,
  \label{eqS:twosided}
\end{equation}
while $A_{kl}-\Lambda_{kl}^2/A_{kl}\ge0$ gives $\beta^2\mathrm{Var}_\rho(V)-Q(V)\ge0$. Together, these
are Eqs.~\eqref{eqS:sandwichgen} and~\eqref{eqS:lemmaQ}. The right-hand inequality
in Eq.~\eqref{eqS:sandwichgen} is an equality if and only if $p_k=p_l$ whenever $V_{kl}\neq0$, i.e., $[H,V]=0$.
\emph{Step 4 (matrix form).} For real $\vec{u}$ let $V_{\vec{u}}=\sum_iu_i\,\partial_iH$. By bilinearity
of Eq.~\eqref{eqS:spectral} in the matrix elements, $\vec{u}^{\,T}Q\,\vec{u}=Q(V_{\vec{u}})$ and
$\vec{u}^{\,T}\Gamma\,\vec{u}=\mathrm{Var}_\rho(V_{\vec{u}})$, so that Eq.~\eqref{eqS:exactQ} reads
\begin{equation}
  \vec{u}^{\,T}Q\,\vec{u}
  =\beta^2\,\mathrm{Var}_\rho(V_{\vec{u}})
  -\beta^3\sum_{k,l}\big|(V_{\vec{u}})_{kl}\big|^2\,(p_k-p_l)(E_l-E_k)\,h\big[\beta(E_l-E_k)\big],
  \qquad h(\ell)=\frac{g(\ell)}{\ell}\ge0 .
  \label{eqS:quadform}
\end{equation}
Each term of the correction is nonnegative, hence $Q\le\beta^2\Gamma$, with
equality if and only if $[H,\partial_iH]=0$ for all $i$. Likewise $\vec{u}^{\,T}K_n\vec{u}=K_n(V_{\vec{u}})$,
so applying Eqs.~\eqref{eqS:sandwichgen} and~\eqref{eqS:lemmaQ} to $V_{\vec{u}}$ gives Eq.~\eqref{eqS:matrix}.
For odd $n$, the matrices $K_n$ are real, symmetric and positive semidefinite: by Eq.~\eqref{eqS:sumrule},
$(K_n)_{ij}=\sum_{kl}(p_k-p_l)(E_l-E_k)^n\,\overline{(\partial_iH)_{kl}}\,(\partial_jH)_{kl}$, with a
nonnegative real weight that is symmetric under $k\leftrightarrow l$. \hfill$\square$

\emph{Relation to the Harris inequality.} The lower bound in Eq.~\eqref{eqS:sandwichgen} is a QFI
analogue of the Harris inequality~\cite{Harris1967}. For Hermitian $V$, Eq.~(20) of~\cite{Harris1967}
reads $\mathrm{Var}_\rho(V)-(\tilde V;\tilde V)\le\frac{\beta}{12}K_1(V)$, where
$(\tilde V;\tilde V)=\sum_{kl}|\tilde V_{kl}|^2\Lambda_{kl}$ is the Bogoliubov (Kubo--Mori) inner product
and $\tilde V=V-\braket{V}_\rho$. Its proof rests on $y\coth y\le1+y^2/3$ [Eq.~(19)
of~\cite{Harris1967}], which is the upper bound in Eq.~\eqref{eqS:abounds} with $y=\ell/2$. Since
$A-\Lambda^2/A=2(A-\Lambda)-(A-\Lambda)^2/A\le2(A-\Lambda)$, Eq.~\eqref{eqS:spectral} and the Harris
inequality give the lower bound in Eq.~\eqref{eqS:sandwichgen}, with the same constant $\beta^3/6$.
Lemma~1 adds the explicit remainder in Eq.~\eqref{eqS:lemmaQ}, which is needed in
Sec.~\ref{app:optimalsystemtwoparthermal}.

\paragraph*{Corollary 1 (exact saturation).}
For a Gibbs state at finite temperature, which has full rank, $Q_{ii}=\beta^2\mathrm{Var}_\rho(\partial_iH)$ for all $i=1,\dots,d$ if and only if
$[H,\partial_iH]=0$ for all $i$. In that case $H$, and hence $\rho$, commutes with every
element of the Lie algebra generated by $\{\partial_iH\}_{i=1}^d$, since by the Jacobi
identity $[H,[A,B]]=[[H,A],B]+[A,[H,B]]$ vanishes whenever $[H,A]=[H,B]=0$. Exact
saturation of the first inequality in Eq.~\eqref{eqS:boundthermal} therefore
requires a Gibbs state invariant under the group generated by this algebra, which also
constrains its fluctuations along elements of the algebra that are not estimated. For
two-component magnetometry this is the obstruction discussed in
Sec.~\ref{app:nogod2}.

\paragraph*{Corollary 2 (asymptotic saturation).}
Let $\tilde V=V-\braket{V}_\rho$. Then, exactly,
\begin{equation}
  1-\frac{Q(V)}{\beta^2\mathrm{Var}_\rho(V)}=\sum_{k,l}\pi_{kl}\,D\big(\beta(E_k-E_l)\big),
  \qquad
  \pi_{kl}=\frac{|\tilde V_{kl}|^2A_{kl}}{\mathrm{Var}_\rho(V)},
  \qquad
  D(y)=1-\frac{\tanh^2(y/2)}{(y/2)^2},
  \label{eqS:deficit}
\end{equation}
where the weights $\pi_{kl}\ge0$ sum to one, and $D$ is even, increasing in $|y|$, and
satisfies $0\le D(y)\le\min\{1,y^2/6\}$. Hence, along any family of Gibbs states,
$Q(V)/[\beta^2\mathrm{Var}_\rho(V)]\to1$ if and only if the weight carried by the pairs
of levels with $\beta|E_k-E_l|>\epsilon$ vanishes for every $\epsilon>0$. A sufficient
condition, which is easy to evaluate, is
\begin{equation}
  \frac{\beta K_1(V)}{\mathrm{Var}_\rho(V)}\;\to\;0,
  \label{eqS:asymcriterion}
\end{equation}
since $1-Q(V)/[\beta^2\mathrm{Var}_\rho(V)]\le\beta K_1(V)/[6\,\mathrm{Var}_\rho(V)]$
by Eq.~\eqref{eqS:sandwichgen}.

\emph{Proof.} In the parametrization used in the proof of Lemma~1, $\Lambda/A=\tanh(\ell/2)/(\ell/2)$,
so that $A-\Lambda^2/A=A\,D(\ell)$. Since $D(0)=0$, the diagonal entries of $V$ do not
contribute to Eq.~\eqref{eqS:spectral}, and $V$ can be replaced by $\tilde V$ there;
dividing by $\beta^2\mathrm{Var}_\rho(V)=\beta^2\sum_{kl}|\tilde V_{kl}|^2A_{kl}$ gives
Eq.~\eqref{eqS:deficit}. $D$ increases with $|y|$ because $\tanh t/t$ decreases
with $|t|$, and $D(y)\le\frac{y}{3}\tanh\frac{y}{2}\le\frac{y^2}{6}$ follows from the upper
bound in Eq.~\eqref{eqS:kernel} together with $(p-q)/A=2\tanh(\ell/2)$. Finally, for any
$\epsilon>0$, $D(\epsilon)\,\pi(\epsilon)\le1-Q(V)/[\beta^2\mathrm{Var}_\rho(V)]\le D(\epsilon)+\pi(\epsilon)$,
where $\pi(\epsilon)$ is the weight of the pairs with $\beta|E_k-E_l|>\epsilon$. \hfill$\square$

\paragraph*{Consequence for the WMSE.}
Since matrix inversion reverses the order between positive definite matrices,
Eq.~\eqref{eqS:matrix} gives, for any probe with $\Gamma-\beta K_1/6>0$,
\begin{equation}
  \frac{1}{\beta^2}\mathrm{Tr}[W\Gamma^{-1}]\;\le\;\mathrm{Tr}[WQ^{-1}]\;\le\;\frac{1}{\beta^2}\mathrm{Tr}\Big[W\Big(\Gamma-\frac{\beta}{6}K_1\Big)^{-1}\Big],
  \label{eqS:WMSEsandwich}
\end{equation}
where $K_1$ is the matrix denoted by $K$ in the main text. The thermal bound is therefore
approached whenever $\beta K_1\ll\Gamma$ in the matrix sense, and saturated if and only if
$K_1=0$.

\subsection{General bounds in multiparameter GS metrology}\label{app:boundGS}
We finally focus on GS metrology. For any real vector $\vec u$ and $V_{\vec u}=\sum_iu_i\partial_iH_{\vec{\theta}}$,
Eq.~\eqref{eqS:QFIMGS} gives, up to $\mathcal{O}(\beta^2e^{-\beta\Delta})$,
\begin{equation*}
  \vec u^{\,T}Q\,\vec u=\sum_{k>0}\frac{4|\braket{k|V_{\vec u}|g_{\vec\theta}}|^2}{(E_k-E_0)^2}
  \le\frac{4}{\Delta^2}\sum_{k>0}|\braket{k|V_{\vec u}|g_{\vec\theta}}|^2
  =\frac{4}{\Delta^2}\mathrm{Var}_{\ket{g_{\vec{\theta}}}}(V_{\vec u}),
\end{equation*}
where we used $E_k-E_0\ge\Delta$ and $\mathrm{Var}_{\ket{g_{\vec{\theta}}}}(V)=\sum_{k>0}|\braket{k|V|g_{\vec{\theta}}}|^2$.
Hence $Q\le\frac{4}{\Delta^2}\Gamma$ with $\Gamma_{ij}=\mathrm{Cov}_{\ket{g_{\vec{\theta}}}}(\partial_iH_{\vec{\theta}},\partial_jH_{\vec{\theta}})$,
which is Eq.~\eqref{eqS:QleqGamma} with $\kappa=4/\Delta^2$. Equality holds if and only if, for all $i$,
$\partial_iH_{\vec{\theta}}\ket{g_{\vec{\theta}}}$ has no component on the excited states with $E_k-E_0\neq\Delta$.
Combining it with Eqs.~\eqref{eqS:matrixchain} and~\eqref{eq:boundvariance},
\begin{align}
    \mathrm{Tr}[WQ^{-1}] \geq \frac{\Delta^2}{4}\mathrm{Tr}[W\Gamma^{-1}]
    \geq \sum_{i=1}^d \frac{w_i \Delta^2}{4\,\mathrm{Var}_{\ket{g_{\vec{\theta}}}}(\partial_i H_{\vec{\theta}})}
    \geq \min_{H^C} \sum_{i=1}^d \frac{w_i \Delta^2}{4\,\mathrm{Var}_{\ket{g_{\vec{\theta}}}}(\partial_i H_{\vec{\theta}})}
    \geq \sum_{i=1}^d \frac{w_i \Delta^2}{\lVert \partial_i H_{\vec{\theta}} \rVert^2}. \label{eq:inequalitiesGSmagnetometry}
\end{align}
The first inequality is saturated under the condition above, the second one whenever $\Gamma$ is diagonal,
and the last one only if the ground state maximizes all the variances simultaneously
(Sec.~\ref{app:upperboundvariance}).

\section{Fundamental bounds for vector magnetometry}\label{app:boundsmagnetometry}
In this section we derive the vector magnetometry bounds reported in the main text. As in the main text, $W=\mathrm{diag}(w_1,\dots,w_d)$ with $\sum_{i=1}^d\sqrt{w_i}=d$.

\subsection{Fundamental bounds in unitary vector magnetometry}\label{app:fundamentalboundsunitaryvectormagnetometry}
Let us first consider unitary vector magnetometry, where we need to minimize the quantity
\begin{equation*}
    \sum_{i=1}^d \frac{w_i}{4\,\mathrm{Var}_{\rho_0} (\mathcal{H}_i)}.
\end{equation*}

We first observe that, by applying the method of Lagrange multipliers,
\begin{equation*}
    \min_{\left\{\rho_0\in \mathcal{H}:\, \sum_{i=1}^d \mathrm{Var}_{\rho_0}(\mathcal{H}_i)=C\right\}}
    \sum_{i=1}^d \frac{w_i}{4\,\mathrm{Var}_{\rho_0}(\mathcal{H}_i)}
    =
    \frac{d^2}{4C}.
\end{equation*}

We now maximize the quantity \(\sum_{i=1}^d \mathrm{Var}_{\rho_0}(\mathcal{H}_i)\). From Eq.~\eqref{eq:boundvariancerotatedderivative}, it follows that
\begin{align*}
    \max_{\rho_0}\sum_{i=1}^d \mathrm{Var}_{\rho_0}(\mathcal{H}_i)
    \leq
    \max_{\rho_0}
    \int_0^1 ds\, \frac{t^2}{\hbar^2}
    \sum_{i=1}^d \mathrm{Var}_{\rho_0}(H'^{{(s)}}_i)\leq
    \frac{t^2}{\hbar^2}
    \max_{\rho_0}
    \max_s
    \sum_{i=1}^d \mathrm{Var}_{\rho_0}(H'^{{(s)}}_i).
\end{align*}

Since \(H'^{{(s)}}_i\) is unitarily equivalent to \(\partial_i H_{\vec{\theta}}\), and the maximization is performed over all states, we have
\begin{equation*}
    \max_{\rho_0}
    \sum_{i=1}^d
    \mathrm{Var}_{\rho_0}(H'^{{(s)}}_i)
    =
    \max_{\rho_0}
    \sum_{i=1}^d
    \mathrm{Var}_{\rho_0}(\partial_i H_{\vec{\theta}}).
\end{equation*}

Therefore,
\begin{equation*}
    \max_{\rho_0}
    \sum_{i=1}^d
    \mathrm{Var}_{\rho_0}(\mathcal{H}_i)
    \leq
    \frac{t^2}{\hbar^2}
    \max_{\rho_0}
    \sum_{i=1}^d
    \mathrm{Var}_{\rho_0}(\partial_i H_{\vec{\theta}})
    \leq
    \frac{t^2 N(N+2)}{4\hbar^2}.
\end{equation*}

Hence,
\begin{equation*}
    \min_{\rho_0}
    \sum_{i=1}^d
    \frac{w_i}{4\,\mathrm{Var}_{\rho_0}(\mathcal{H}_i)}
    =
    \min_{\left\{\rho_0\in\mathcal{H}:\,
    \sum_{i=1}^d \mathrm{Var}_{\rho_0}(\mathcal{H}_i)
    \leq
    \frac{t^2 N(N+2)}{4\hbar^2}\right\}}
    \sum_{i=1}^d
    \frac{w_i}{4\,\mathrm{Var}_{\rho_0}(\mathcal{H}_i)}
    \geq
    \frac{d^2\hbar^2}{t^2 N(N+2)}.
\end{equation*}

Substituting this result into Eq.~\eqref{eqS:boundunitary} yields the bound for unitary vector magnetometry,
\begin{align}
    \mathrm{Tr}[WQ^{-1}]
    \geq \min_{\rho_0}\sum_{i=1}^d \frac{w_i}{4\,\mathrm{Var}_{\rho_0} (\mathcal{H}_i)}
    \geq \frac{d^2\hbar^2}{t^2 N(N+2)}. \label{eqS:boundmagnetometryunitary}
\end{align}

\subsection{Fundamental bounds in thermal equilibrium and GS vector magnetometry}
Let us now consider thermal equilibrium and GS vector magnetometry. Unlike the unitary case, here we need to directly minimize the quantity
\begin{equation*}
    \sum_{i=1}^d \frac{w_i}{\mathrm{Var}_{\rho}(J_i)}.
\end{equation*}

Using Lagrange multipliers, one finds
\[
    \min_{\rho\in\mathcal{H},\,\sum_{i=1}^d \mathrm{Var}_\rho(J_i) = C} \sum_{i=1}^d \frac{w_i}{\mathrm{Var}_\rho(J_i)} = \frac{d^2}{C},
\]
which is attained for $\mathrm{Var}_\rho(J_i) = \sqrt{w_i}\, C/d$. We now observe that
\begin{align*}
    \sum_{i=1}^3 \braket{J_i^2}_\rho = \braket{\vec{J}^2}_\rho \leq \frac{N(N+2)}{4},
\end{align*}
where we upper bounded the averaged $\vec{J}^2$ by its largest eigenvalue. Consequently, $\sum_{i=1}^d \mathrm{Var}_{\rho}(J_i) \leq \frac{N(N+2)}{4}$, and therefore
\begin{align}
    \sum_{i=1}^d \frac{w_i}{ \mathrm{Var}_{\rho}(J_i)}
    \geq \frac{4d^2}{N(N+2)}. \label{eq:boundsuminversevariancesangularmomentum}
\end{align}

The saturation of the bound requires a state satisfying
\begin{equation}
\mathrm{Var}_\rho(J_i) = \frac{\sqrt{w_i}N(N+2)}{4d},
\label{eq:optimalvariances}
\end{equation}
for all \(d\) directions spanning the magnetic field, while \(\mathrm{Var}_\rho(J_i)=0\) for the remaining directions.
By applying Eq.~\eqref{eq:boundsuminversevariancesangularmomentum} to Eqs.~\eqref{eqS:boundthermal} and \eqref{eq:inequalitiesGSmagnetometry}, we obtain the corresponding bounds for thermal equilibrium and for ground states, respectively:
\begin{align}
    \mathrm{Tr}[WQ^{-1}]
    &\geq \min_{H^C} \sum_{i=1}^d \frac{w_i}{\beta^2 \mathrm{Var}_{\rho}(J_i)}
    \geq \frac{4d^2}{\beta^2 N(N+2)}, \label{eqS:boundmagnetometrythermal}\\
    \mathrm{Tr}[WQ^{-1}]
    &\geq \min_{H^C} \sum_{i=1}^d \frac{w_i \Delta^2}{4\,\mathrm{Var}_{\rho}(J_i)}
    \geq \frac{\Delta^2 d^2}{N(N+2)}. \label{eqS:boundmagnetometryGS}
\end{align}

\subsection{\texorpdfstring{No-go for exact saturation in two-component thermal magnetometry}{No-go for exact saturation in two-component thermal magnetometry}}\label{app:nogod2}
Section~\ref{app:optimalsystems} shows that, for $W=\mathbb{I}_d$, the bounds above are saturated
for $d=3$ in all three settings and for $d=2$ in the unitary and GS settings. In two-component
thermal magnetometry, instead, they cannot be saturated exactly.

\paragraph*{Proposition 1.} Let $H_{\vec\theta}=\theta_xJ_x+\theta_yJ_y+H^C$ and let
$\rho$ be its Gibbs state at the working point. (i) No control Hamiltonian $H^C$
saturates $\mathrm{Tr}[Q^{-1}]\ge16/[\beta^2N(N+2)]$. (ii) If $H_{\vec\theta}$ commutes
with $J_x$ and $J_y$ at the working point, which by Corollary~1 is exactly the case in
which the first inequality in Eq.~\eqref{eqS:boundthermal} is saturated, then
\begin{equation}
  \mathrm{Tr}[Q^{-1}]\;\ge\;\frac{24}{\beta^2N(N+2)}=\frac32\cdot\frac{16}{\beta^2N(N+2)},
  \label{eqS:nogo}
\end{equation}
with equality approached for $H^C=-MJ^2$ as $\beta M\to\infty$.

\emph{Proof.} If $[H_{\vec\theta},J_x]=[H_{\vec\theta},J_y]=0$, the Jacobi identity
gives $[H_{\vec\theta},J_z]=-i[H_{\vec\theta},[J_x,J_y]]=0$, so $\rho$ commutes with all
three components and is invariant under all rotations $e^{-i\phi\,\vec n\cdot\vec J}$.
Hence $\braket{J_i}=0$, $\braket{J_x^2}=\braket{J_y^2}=\braket{J_z^2}=\braket{\vec J^2}/3$,
and $\mathrm{Cov}_\rho(J_x,J_y)=0$ (a rotation by $\pi/2$ about $z$ maps it to minus itself).
By Theorem~1, $Q=\beta^2\Gamma=\beta^2\braket{\vec J^2}\mathbb{I}_2/3$, so that
$\mathrm{Tr}[Q^{-1}]=6/(\beta^2\braket{\vec J^2})\ge24/[\beta^2N(N+2)]$, with equality if
and only if $\rho$ is supported on the sector $j=N/2$; this is approached by $H^C=-MJ^2$ as
$\beta M\to\infty$ (Sec.~\ref{app:optimalsystemthreeparthermal}). This proves (ii). For (i),
saturation of the bound requires equality in every step of the chain leading to it, in
particular $Q_{ii}=\beta^2\mathrm{Var}_\rho(J_i)$ for $i=x,y$, which by Corollary~1 means
$[H_{\vec\theta},J_x]=[H_{\vec\theta},J_y]=0$; then (ii) applies and the bound is
missed by a factor $3/2$. \hfill$\square$

\section{Optimal systems and optimal measurements in vector magnetometry}\label{app:optimalsystems}
In this section, we prove the optimality of the systems presented in the main text for vector magnetometry. Moreover,
we show that all these systems satisfy the WCC, meaning that the QCRB is
asymptotically attainable.

Furthermore, for unitary and ground-state metrology, we also present the
optimal projective measurements. Indeed, when $\rho_{\vec{\theta}}$ is pure,
the WCC reduces to
\begin{equation}
    \mathrm{Im}\!\left( \braket{\partial_i \psi_{\vec{\theta}}
    | \partial_j \psi_{\vec{\theta}}} \right) = 0,
    \quad \forall\, i,j = 1,\ldots,d,
    \label{eq:WCCpurestate}
\end{equation}
and, when this condition holds, the QCRB can be asymptotically attained with
projective measurements. More precisely, Ref.~\cite{Pezze_2017} derives
theorems on optimal projective measurements under these conditions. In
particular, given a pure state $\ket{\psi_{\vec{\theta}}}$ and a set of
projectors $\{\ketbra{m_k}{m_k}\}$ such that $\ket{m_0} =
\ket{\psi_{\vec{\theta}}}$ and $\braket{m_k|\psi_{\vec{\theta}}} = 0$ for
$k \neq 0$, the classical and quantum Fisher information matrices coincide,
$F(\vec{\theta}) = Q(\vec{\theta})$, if and only if
\begin{equation}
    \mathrm{Im}\Big[ \braket{\partial_i \psi_{\vec{\theta}} | m_k}
    \braket{m_k | \partial_j \psi_{\vec{\theta}}} \Big] = 0
    \quad \forall\, i,j = 1,\ldots,d \text{ and } \forall\, k \neq 0.
    \label{eq:purestateoptmeasorth}
\end{equation}
These conditions make it straightforward to construct the optimal projective
measurements, which we also present in the following.

\subsection{\texorpdfstring{Optimal system for three-parameter estimation with unitary dynamics}{Optimal system for three-parameter estimation with unitary dynamics}}\label{app:optimalsystemthreeparunitary}
For $H^C=0$ and local estimation around $\vec\theta=0$, the generators are $\mathcal{H}_i=tJ_i$.
Consider the XYZ-GHZ state of the main text,
\begin{equation}
    \ket{\psi} \coloneqq \mathcal{N} (\ket{\Phi_1}+\ket{\Phi_2}+\ket{\Phi_3}), \label{eq:3parGHZstate}
\end{equation}
with $\ket{\Phi_k}=(\ket{\phi^+_k}^{\otimes N}+\ket{\phi^-_k}^{\otimes N})/\sqrt2$. For $N=8n$,
Proposition~2 (Sec.~\ref{app:variances}) gives $\braket{J_i}=0$ and
$\Gamma=t^2\frac{N(N+2)}{12}\mathbb{I}_3$. Since the state is pure, $Q=4\Gamma$ and
$\mathrm{Tr}[Q^{-1}]=9/[t^2N(N+2)]$, which saturates Eq.~\eqref{eqS:boundmagnetometryunitary}
for $W=\mathbb{I}_3$. The WCC~\eqref{eq:WCCpurestate} holds, since
$\mathrm{Im}\braket{\partial_i\psi|\partial_j\psi}=t^2\,\mathrm{Im}\braket{J_iJ_j}
=\frac{t^2}{2}\sum_k\epsilon_{ijk}\braket{J_k}=0$. Optimal measurements are given
in~\cite{Baumgratz_2016}.

\subsection{Optimal system for two-parameter estimation with unitary dynamics} \label{app:optimalsystemtwoparunitary}
Let us next consider two-parameter estimation under unitary dynamics, namely the estimation of the components $\theta_1,\theta_2$ of a magnetic field along the x and y axis, respectively. We now show that the Dicke state $\ket{\psi}=\ket{N/2,0}$ is optimal, as it saturates the bound in Eq.~\eqref{eqS:boundmagnetometryunitary} for $d=2$.

To proceed, we recall that
\begin{align*}
    J_1 \ket{\tfrac{N}{2},0}
&= \frac{1}{2}\sqrt{\tfrac{N}{2}\!\left(\tfrac{N}{2}+1\right)}
\left(\ket{\tfrac{N}{2},1} + \ket{\tfrac{N}{2},-1}\right),\\
J_2 \ket{\tfrac{N}{2},0}
&= \frac{1}{2i}\sqrt{\tfrac{N}{2}\!\left(\tfrac{N}{2}+1\right)}
\left(\ket{\tfrac{N}{2},1} - \ket{\tfrac{N}{2},-1}\right).
\end{align*}

From this, and using Eq.~\eqref{eq:QFIMunitary}, we obtain
\begin{align*}
    Q_{11} &= 4t^2 \mathrm{Var}_{\ket{\psi}}(J_1)
    = \frac{N(N+2)t^2}{2}
    = Q_{22},\\
    Q_{12} &= 4t^2 \mathrm{Cov}_{\ket{\psi}}(J_1,J_2) = 0,
\end{align*}
which indeed saturates Eq.~\eqref{eqS:boundmagnetometryunitary} for $d=2$.

We now show that the WCC (Eq.~\eqref{eq:WCCpurestate}) is satisfied. The derivatives of $\ket{\psi}$ can be calculated by expanding the state over small variations of the parameters $\theta_1, \theta_2$:
\begin{align*}
    \ket{\partial_1 \psi}
&= \lim_{\theta_1 \rightarrow 0}
\frac{e^{-i\theta_1 J_1 t}\ket{\tfrac{N}{2},0}-\ket{\tfrac{N}{2},0}}{\theta_1}
= -it J_1 \ket{\tfrac{N}{2},0} = -\frac{it}{2}\sqrt{\tfrac{N}{2}\!\left(\tfrac{N}{2}+1\right)}
\left(\ket{\tfrac{N}{2},1} + \ket{\tfrac{N}{2},-1}\right),\\[6pt]
\ket{\partial_2 \psi}
&= \lim_{\theta_2 \rightarrow 0}
\frac{e^{-i\theta_2 J_2 t}\ket{\tfrac{N}{2},0}-\ket{\tfrac{N}{2},0}}{\theta_2}
= -it J_2 \ket{\tfrac{N}{2},0} = -\frac{t}{2}\sqrt{\tfrac{N}{2}\!\left(\tfrac{N}{2}+1\right)}
\left(\ket{\tfrac{N}{2},1} - \ket{\tfrac{N}{2},-1}\right).
\end{align*}

The two derivatives are orthogonal, thus Eq.~\eqref{eq:WCCpurestate} is satisfied.

\subsubsection{Optimal measurement for two-parameter estimation with unitary dynamics}\label{app:pureWCCoptimalmeas}
Since the parametrized state is pure and the WCC is satisfied, the QCRB can be asymptotically saturated by applying the same projective measurement independently on each repetition of the experiment. 

Such a projective measurement can then be constructed as
\begin{align*}
    \ket{m_0} &\coloneqq \ket{\psi}, \\
    \ket{m_1} &\coloneqq \frac{\ket{\partial_1\psi}}{\sqrt{\braket{\partial_1\psi|\partial_1\psi}}}, \\
    \ket{m_2} &\coloneqq \frac{\ket{\partial_2\psi}}{\sqrt{\braket{\partial_2\psi|\partial_2\psi}}},
\end{align*}
and completed to an orthonormal basis with arbitrary orthogonal vectors.

\subsection{Optimal system for three-parameter estimation at thermal equilibrium}
\label{app:optimalsystemthreeparthermal}

Let us now consider three-parameter estimation at thermal equilibrium. Given the control Hamiltonian $H^C = -M J^2$, the full Hamiltonian reads
\[
    H_{\vec{\theta}} = \sum_{i=1}^3 \theta_i J_i - M J^2.
\]
The corresponding Gibbs state around $\theta_i = 0$, in the limit $M \gg \beta^{-1}$, is
\begin{equation}
    \rho_{\vec{0}}
    =
    \frac{1}{N+1}
    \sum_{m=-\frac{N}{2}}^{\frac{N}{2}}
    \ket{\frac{N}{2},m}\bra{\frac{N}{2},m}. \label{eq:maximallymixedstatesymmetricsubspace}
\end{equation}
Since $[H_{\vec{\theta}},J_i]=0$ at $\vec\theta=0$ for all $i$, Theorem~1 of Sec.~\ref{app:thermalsandwich} gives
$Q=\beta^2\Gamma$, with $\Gamma_{ij}=\mathrm{Cov}_\rho(J_i,J_j)$. The state~\eqref{eq:maximallymixedstatesymmetricsubspace}
is the maximally mixed state of the spin-$j$ irreducible representation, $j=N/2$, and is therefore invariant
under all rotations. Hence $\braket{J_i}=0$ and $\frac12\braket{\{J_i,J_j\}}=\delta_{ij}\braket{\vec J^2}/3=\delta_{ij}\,j(j+1)/3$,
so that
\begin{equation}
  Q=\beta^2\,\frac{N(N+2)}{12}\,\mathbb{I}_3,
\end{equation}
which saturates the three-parameter thermal bound for $W=\mathbb{I}_3$.

The WCC also follows from symmetry. Since $\rho$ commutes with every $J_k$, $\mathbb{J}_{L,\rho}[J_k]=\rho J_k$, so
that $\partial_k\rho=-\beta\rho\,(J_k-\braket{J_k})$ and the SLDs are $L_k=-\beta(J_k-\braket{J_k})=-\beta J_k$. Therefore
\[
  \mathrm{Tr}\big(\rho[L_l,L_m]\big)=\beta^2\,\mathrm{Tr}\big(\rho[J_l,J_m]\big)=i\beta^2\sum_n\epsilon_{lmn}\braket{J_n}=0 .
\]
These two arguments hold for any $M$, since $H^C=-MJ^2$ commutes with all $J_k$; the limit $\beta M\gg1$ is
only needed to populate the maximal-spin sector, where $\braket{\vec J^2}=N(N+2)/4$ is maximal.

\subsection{Optimal system for two-parameter estimation at thermal equilibrium}
\label{app:optimalsystemtwoparthermal}

For $d=2$ thermal magnetometry, the bound $\mathrm{Tr}[WQ^{-1}]\ge
4d^2/[\beta^2N(N+2)]$ cannot be saturated exactly: by Corollary~1 of
Sec.~\ref{app:thermalsandwich}, exact saturation would require a rotationally invariant
Gibbs state, which spends a third of the variance budget on $J_z$
(Sec.~\ref{app:nogod2}). The bound nevertheless gives the correct asymptotic
precision: by Corollary~2, together with $\braket{J_z^2}_\rho\ll N^2$, it is attained in the limit $N\to\infty$ by the whole
family of control Hamiltonians
\begin{equation}
  H^C = c\,J_z^2 - M\,J^2 ,
  \qquad \beta M \gg 1,
  \qquad \beta c \propto N^{-\gamma}
  \quad\text{with}\quad 0<\gamma<2 ,
  \label{eqS:HCthermald2}
\end{equation}
the fastest convergence being obtained for $\beta c=\sqrt{6/[N(N+2)]}$, for
which
\begin{equation}
  \mathrm{Tr}[Q^{-1}] = \frac{16}{\beta^2 N(N+2)}
  \left[\,1+\frac{4}{\sqrt{6N(N+2)}}+\mathcal{O}(N^{-2})\,\right].
  \label{eqS:resultd2}
\end{equation}
Here $Q$ denotes the QFIM of the two estimated components $\theta_x,\theta_y$; below we
also compute the full $3\times3$ QFIM, obtained by adding a term $\theta_zJ_z$ to the
Hamiltonian.

By Proposition~1, exact saturation requires $\rho$ to commute with $J_x$ and $J_y$ and, at the same
time, $\braket{J_z^2}=0$, two conditions that cannot hold together. Both can, however,
be approached simultaneously: if the coupling is weak, $x\equiv\beta c\to0$,
neighboring populations are nearly equal and $\rho$ almost commutes with
$J_{x,y}$, while if $xN^2\to\infty$ only a small fraction of the levels is populated
and $\braket{J_z^2}\ll N^2$. We quantify this with Theorem~1 and Corollary~2 of
Sec.~\ref{app:thermalsandwich}.

The proof of Eq.~\eqref{eqS:resultd2} is organized as follows: the QFIM is diagonal by
symmetry (Sec.~\ref{app:diagd2}), Theorem~1 gives two-sided bounds on $Q_{xx}$
(Sec.~\ref{app:transversed2}), and these bounds are evaluated for $\beta c\propto N^{-\gamma}$
and optimized over $\beta c$ (Sec.~\ref{app:scalingsd2}). Section~\ref{app:longitudinald2}
computes the longitudinal component $Q_{zz}$, which is not needed for
Eq.~\eqref{eqS:resultd2}. Finally, Sec.~\ref{app:WCCd2} shows that the WCC holds exactly,
so that the QCRB is asymptotically attainable.

\subsubsection{Diagonality of the QFIM}\label{app:diagd2}

We now turn to the family~\eqref{eqS:HCthermald2} and compute its QFIM, including the
longitudinal component $\theta_z$. Since $[H^C,J_z]=0$, the rotations
$R_\phi=e^{-i\phi J_z}$ leave $H^C$ invariant and rotate the signal,
$R_\phi(\vec\theta\cdot\vec J)R_\phi^\dagger=(O_\phi\vec\theta)\cdot\vec J$, where $O_\phi$
is the rotation by the angle $\phi$ about the $z$ axis. Hence
$R_\phi\,\rho_{\vec\theta}\,R_\phi^\dagger=\rho_{O_\phi\vec\theta}$ and, since the QFIM is
invariant under unitaries that do not depend on $\vec\theta$, the QFIM at the working point
$\vec\theta=0$ obeys $Q=O_\phi^TQ\,O_\phi$ for every $\phi$. A symmetric matrix with this
property has an isotropic $x$--$y$ block and vanishing $Q_{xz}$, $Q_{yz}$, so that
\begin{equation}
  Q=\mathrm{diag}\left(Q_{xx},\,Q_{xx},\,Q_{zz}\right)
  \label{eqS:Qdiagd2}
\end{equation}
exactly, for every $N$, $\beta$, $c$ and $M$. The same argument applies to $\Gamma$ and to the
matrix $K_1$ of Theorem~1, since $\rho$ and $H$ are invariant and the $J_a$ transform as a
vector. In particular, the second inequality in Eq.~\eqref{eqS:boundthermal}, which requires a
diagonal $\Gamma$, is satisfied for free, and it remains to compute $Q_{xx}$ and $Q_{zz}$.

\subsubsection{The transverse component}\label{app:transversed2}

For $\beta M\gg1$ the Gibbs state is supported on the maximal-spin sector $j=N/2$ (the
populations of the other sectors are exponentially small in $\beta M$), where
$J^2=A_0\equiv j(j+1)=N(N+2)/4$. Since the $J_a$ do not couple different sectors, we can work
on that sector. Writing $\ket{m}\equiv\ket{j,m}$ and $x\equiv\beta c$,
\begin{equation}
  \rho=\sum_{m=-j}^{j} p_m \ket{m}\bra{m},
  \qquad
  p_m=\frac{e^{-x m^2}}{Z},
  \qquad
  Z=\sum_{m=-j}^{j}e^{-x m^2}.
  \label{eqS:rhod2}
\end{equation}
In this basis, with $C_m\equiv\sqrt{A_0-m(m+1)}>0$,
\begin{equation}
  (J_x)_{m+1,m}=(J_x)_{m,m+1}=\frac{C_m}{2},\qquad
  (J_y)_{m+1,m}=-(J_y)_{m,m+1}=\frac{C_m}{2i},\qquad
  (J_z)_{mm}=m,
  \label{eqS:matel}
\end{equation}
all other entries vanishing. Thus $J_x$ and $J_y$ only connect neighboring levels $m$ and
$m+1$, whose energies differ by $E_{m+1}-E_m=c(2m+1)$. In particular
$\braket{J_x}=\braket{J_y}=0$, and $\braket{J_z}=\sum_mp_m\,m=0$ since $p_m$ is even in $m$.
We now apply Theorem~1 of Sec.~\ref{app:thermalsandwich} to $V=J_x$ and $H=cJ_z^2-MJ^2$,
and compute the three ingredients of Eq.~\eqref{eqS:lemmaQ} in turn.

\emph{Variance.} By Eq.~\eqref{eqS:Qdiagd2} applied to $\Gamma$,
$\braket{J_x^2}=\braket{J_y^2}$, while $J_x^2+J_y^2=A_0-J_z^2$ on the sector. Hence
$\mathrm{Var}_\rho(J_x)=\braket{J_x^2}=(A_0-\braket{m^2})/2$.

\emph{Double commutator.} The term $-MJ^2$ commutes with $J_x$, while
$[H,J_x]=c[J_z^2,J_x]=ic\{J_z,J_y\}$ and $[J_x,[H,J_x]]=2c\,(J_y^2-J_z^2)$ (expand
$[J_x,J_zJ_y+J_yJ_z]$ with $[J_x,J_y]=iJ_z$ and $[J_x,J_z]=-iJ_y$). Therefore
\begin{equation*}
  K_1(J_x)=\braket{[J_x,[H,J_x]]}=2c\big(\braket{J_y^2}-\braket{J_z^2}\big)=c\,\big(A_0-3\braket{m^2}\big).
\end{equation*}

\emph{Remainder.} With $u\equiv m+\frac12$, we have $C_m^2=A_0-m(m+1)=\frac{(N+1)^2}{4}-u^2$
and $E_{m+1}-E_m=2cu$. By Eq.~\eqref{eqS:matel} and the sum rule~\eqref{eqS:sumrule},
\begin{equation*}
  K_3(J_x)=\sum_{k,l}|(J_x)_{kl}|^2(p_k-p_l)(E_l-E_k)^3
  =4c^3\!\sum_{m=-j}^{j-1}\!C_m^2\,u^3\,(p_m-p_{m+1})
  \le c^3(N+1)^2\!\sum_{m=-j}^{j-1}\!u^3(p_m-p_{m+1}),
\end{equation*}
since every term is nonnegative and $C_m^2\le(N+1)^2/4$. Summation by parts, with
$(m+\tfrac12)^3-(m-\tfrac12)^3=3m^2+\tfrac14$, gives
$\sum_{m=-j}^{j-1}u^3(p_m-p_{m+1})=3\braket{m^2}+\frac14-2(j+\tfrac12)^3p_j\le3\braket{m^2}+\frac14$,
so that $K_3(J_x)\le c^3(N+1)^2\big(3\braket{m^2}+\frac14\big)$.

\paragraph*{Lemma 2.}
For $p_m\propto e^{-xm^2}$ on $m=-j,-j+1,\dots,j$ and $0<x\le1$,
\begin{equation}
  \frac{1-(N+1)p_j}{2x}-1\;\le\;\braket{m^2}\;\le\;\frac{e^{x}}{2x},
  \qquad p_j\le e^{-xN^2/4}.
  \label{eqS:m2bounds}
\end{equation}
\emph{Proof.} Extend $p_m=e^{-xm^2}/Z$ to all indices $m$, with the same $Z$. Then
$\ln(p_{m-1}/p_{m+1})=4xm$ and $\sqrt{p_{m-1}p_{m+1}}=e^{-x}p_m$, so that
$p_{m-1}-p_{m+1}=4xm\,\Lambda(p_{m-1},p_{m+1})$, with
$e^{-x}p_m\le\Lambda(p_{m-1},p_{m+1})\le\frac12(p_{m-1}+p_{m+1})$ (the logarithmic
mean lies between the geometric and arithmetic means). Summation by parts, with
$p_{-m}=p_m$, gives $\sum_{m=-j}^{j}m\,(p_{m-1}-p_{m+1})=2-2(j+1)p_j-2jp_{j+1}$, which
lies between $2-2(N+1)p_j$ and $2$ since $p_{j+1}\le p_j$. The lower bound on $\Lambda$
yields $4xe^{-x}\braket{m^2}\le2$; the upper bound yields
$2-2(N+1)p_j\le2x\sum_{|m|\le j}m^2(p_{m-1}+p_{m+1})\le2x(2\braket{m^2}+2)$, because
$\sum_{|m|\le j}m^2(p_{m-1}+p_{m+1})=\sum_{|k|\le j}[(k+1)^2+(k-1)^2]p_k-2(j+1)^2p_j+2j^2p_{j+1}$.
Finally, $Z\ge1$ for $0<x\le1$ (for integer $j$ the term $m=0$ already gives $1$; for
half-integer $j$, $Z\ge2e^{-x/4}\ge1$), so that $p_j\le e^{-xj^2}=e^{-xN^2/4}$. \hfill$\square$

\paragraph*{Result for the transverse component.}
Inserting the three ingredients into Eq.~\eqref{eqS:lemmaQ},
\begin{equation}
  0\;\le\;\frac{Q_{xx}}{\beta^2}-\frac{A_0\big(1-\frac x3\big)-(1-x)\braket{m^2}}{2}
  \;\le\;\frac{x^3(N+1)^2}{96}\Big(3\braket{m^2}+\frac14\Big),
  \label{eqS:sandwich}
\end{equation}
for every $N$ and every $x>0$. Here $x/3$ is the loss due to the finite level spacing,
and $\braket{m^2}/A_0$ the fraction of the variance budget assigned to the $z$ direction.
For $0<x\le1$, Lemma~2 turns Eq.~\eqref{eqS:sandwich} into explicit two-sided bounds;
the lower one is
\begin{equation}
\frac{2\,Q_{xx}}{\beta^2A_0}\;\ge\;1-\frac{x}{3}-\frac{(1-x)\,e^{x}}{2xA_0},
  \label{eqS:Qxxfinal}
\end{equation}
and the upper one, using $(N+1)^2/(4A_0)\le4/3$, reads
\begin{equation}
  \frac{2\,Q_{xx}}{\beta^2A_0}\;\le\;1-\frac{x}{3}-\frac{1}{2xA_0}
  +\frac{x^2e^{x}}{6}+\frac{x^3}{36}+\frac{3}{2A_0}+\frac{(N+1)e^{-xN^2/4}}{2xA_0}.
  \label{eqS:Qxxupper}
\end{equation}
Hence, for $x=\gamma_0N^{-\gamma}$ with $0<\gamma<2$, so that $x\to0$ and the last term
in Eq.~\eqref{eqS:Qxxupper} decays faster than any power of $1/N$,
\begin{equation}
  \frac{2\,Q_{xx}}{\beta^2A_0}=1-\frac{x}{3}-\frac{1}{2xA_0}+\mathcal{O}(x^2)+\mathcal{O}(A_0^{-1}).
  \label{eqS:Qxxasym}
\end{equation}
In the language of Corollary~2, $\beta K_1(J_x)/\mathrm{Var}_\rho(J_x)=2x(A_0-3\braket{m^2})/(A_0-\braket{m^2})\simeq2x\to0$.
Since $Q$, $\Gamma$ and $K_1$ are diagonal for this model, the lower bound~\eqref{eqS:Qxxfinal}
is equivalently obtained from Eq.~\eqref{eqS:WMSEsandwich}, which here reads
$\mathrm{Tr}[Q^{-1}]\le2/\{\beta^2[\mathrm{Var}_\rho(J_x)-\beta K_1(J_x)/6]\}$; this is the form
used in the main text.

\subsubsection{The longitudinal component}\label{app:longitudinald2}
Since $J_z$ commutes with the Hamiltonian, $Q_{zz}=\beta^2\mathrm{Var}_\rho(J_z)=\beta^2\braket{m^2}$,
and Lemma~2 gives
\begin{equation}
  Q_{zz}=\frac{\beta^2}{2x}\Big[1+\mathcal{O}(x)+\mathcal{O}\big(N e^{-xN^2/4}\big)\Big].
  \label{eqS:Qzz}
\end{equation}
This component therefore scales as $N^{\gamma}$, in contrast with the $N^2$ scaling of $Q_{xx}$.

\subsubsection{Admissible scalings and the optimal choice}\label{app:scalingsd2}

With $W=\mathbb{I}_2$ and $Q=Q_{xx}\mathbb{I}_2$ on the estimated block,
$\mathrm{Tr}[Q^{-1}]=2/Q_{xx}$, so that
\begin{equation}
  R\equiv\frac{\mathrm{Tr}[Q^{-1}]}{16/[\beta^2N(N+2)]}=\frac{\beta^2A_0}{2\,Q_{xx}}
  =1+\frac{x}{3}+\frac{1}{2xA_0}+\mathcal{O}(x^2)+\mathcal{O}\big((xN^2)^{-2}\big)+\mathcal{O}(N^{-2})
  \label{eqS:R}
\end{equation}
by Eq.~\eqref{eqS:Qxxasym}, while Eq.~\eqref{eqS:Qxxfinal} gives the explicit bound
$R\le[1-\epsilon(x)]^{-1}$ with $\epsilon(x)=\frac x3+\frac{(1-x)e^x}{2xA_0}$, valid for all
$N$ whenever $0<x\le1$ and $\epsilon(x)<1$. Hence $R\to1$ whenever $x\to0$ and
$xN^2\to\infty$. Writing $x=\gamma_0N^{-\gamma}$,
\begin{equation*}
  R-1\;=\;\frac{2}{\gamma_0}N^{\gamma-2}+\frac{\gamma_0}{3}N^{-\gamma}
  +\text{(higher orders)}
  \;\xrightarrow[N\to\infty]{}\;0
  \qquad\text{for all } 0<\gamma<2 .
\end{equation*}
The convergence is fastest at $\gamma=1$, where the two leading corrections balance.
They are equal at
\begin{equation*}
  \beta c^\star=x^\star=\sqrt{\frac{6}{N(N+2)}},
  \qquad
  \frac{x^\star}{3}=\frac{1}{2x^\star A_0}=\frac{2}{\sqrt{6N(N+2)}},
\end{equation*}
so that $R=1+4/\sqrt{6N(N+2)}+\mathcal{O}(N^{-2})$, which proves Eq.~\eqref{eqS:resultd2}.
The same choice gives the explicit bound with
$\epsilon(x^\star)=[1+(1-x^\star)e^{x^\star}]/\sqrt{6A_0}=4/\sqrt{6N(N+2)}+\mathcal{O}(N^{-3})$.
At this point the longitudinal component~\eqref{eqS:Qzz} reads
$Q_{zz}\simeq\beta^2N/(2\sqrt6)$: the probe attains Heisenberg scaling on the two
estimated components at the price of standard linear scaling on the third, spending
only $\mathcal{O}(N)$ of the budget $\sum_i\mathrm{Var}_\rho(J_i)\le N(N+2)/4$ on $J_z$.

\subsubsection{The weak commutativity condition holds exactly}\label{app:WCCd2}

Let $U=e^{-i\pi J_x}$, under which $J_x\mapsto J_x$ while $J_y\mapsto-J_y$ and
$J_z\mapsto-J_z$. Since $H^C=c\,J_z^2-MJ^2$ is even in $J_z$ it is invariant
under $U$, so that the full Hamiltonian obeys
$U\,H(\theta_x,\theta_y,\theta_z)\,U^\dagger=H(\theta_x,-\theta_y,-\theta_z)$.
As conjugation commutes with any function of $H$ and leaves the trace, hence
the partition function, invariant, this lifts to the state:
\begin{equation}
  U\,\rho_{(\theta_x,\theta_y,\theta_z)}\,U^\dagger
  =\rho_{(\theta_x,-\theta_y,-\theta_z)} ,
  \label{eqS:parityfamily}
\end{equation}
an identity in the parameters rather than a statement at a single point.
Evaluating Eq.~\eqref{eqS:parityfamily} at the working point $\vec\theta=0$
gives $U\rho U^\dagger=\rho$, whereas differentiating it there gives, by the
chain rule,
\begin{equation*}
  U\,\partial_x\rho\,U^\dagger=\partial_x\rho ,
  \qquad
  U\,\partial_y\rho\,U^\dagger=-\partial_y\rho .
\end{equation*}
Conjugating the defining relation $\partial_a\rho=\tfrac12(L_a\rho+\rho L_a)$
by $U$ and using $U\rho U^\dagger=\rho$, the operators $UL_xU^\dagger$ and
$-UL_yU^\dagger$ are seen to obey the same two relations as $L_x$ and $L_y$;
since $\rho$ is full rank the SLD is unique, and therefore
$UL_xU^\dagger=L_x$ and $UL_yU^\dagger=-L_y$. The commutator is thus odd,
$U\,[L_x,L_y]\,U^\dagger=-[L_x,L_y]$, and
\begin{equation}
  \mathrm{Tr}\big(\rho\,[L_x,L_y]\big)
  =\mathrm{Tr}\big(U\rho U^\dagger[L_x,L_y]\big)
  =\mathrm{Tr}\big(\rho\,U^\dagger[L_x,L_y]U\big)
  =-\mathrm{Tr}\big(\rho\,[L_x,L_y]\big)=0
\end{equation}
identically, for every $N$ and every $x$. The QCRB is therefore asymptotically
attainable with collective measurements.

\subsection{Optimal system for three-parameter estimation with a ground state}\label{app:optimalsystemthreeparGS}
Here we focus on three-parameter GS magnetometry and consider the following control Hamiltonians.

The first choice,
\begin{equation}
    H_1^C = -\frac{\Delta}{2}\ket{\psi}\bra{\psi}
    + \frac{\Delta}{2}\Big(\sum_{i=1}^{\dim(\mathcal{H})-1}
    \ket{\psi_{\perp,i}}\bra{\psi_{\perp,i}} \Big),
    \label{eq:optimalGShamiltonian1}
\end{equation}
where $\ket{\psi}$ is the XYZ-GHZ state~\eqref{eq:3parGHZstate}
and $\{\ket{\psi_{\perp,i}}\}$ complete an orthonormal basis of $\mathcal{H}$, generates a spectral gap $\Delta$ but leaves the excited subspace fully degenerate.

To reduce this degeneracy, we also introduce
\begin{equation}
    H_2^C = -\frac{\Delta}{2}\ket{\psi}\bra{\psi}
    + \frac{\Delta}{2}\Big(\sum_{i=1}^{3} \ket{\psi_{i}}\bra{\psi_{i}} \Big)
    + \sum_{i=4}^{\dim(\mathcal{H})-1} E_i \ket{\tilde{\psi}_{i}}\bra{\tilde{\psi}_{i}},
    \label{eq:optimalGShamiltonian2}
\end{equation}
where
\[
    \ket{\psi_{i}} =
    \frac{J_i \ket{\psi}}{\sqrt{\braket{\psi |J_i^2| \psi}}}, \quad i=1,2,3,
\]
are normalized states spanning the excited subspace, while $\{\ket{\tilde{\psi}_{i}}\}$ complete an orthonormal basis orthogonal to both $\ket{\psi}$ and $\{\ket{\psi_i}\}$, with $E_i - E_0 \geq \Delta$ for $i>3$.

We now show that these control Hamiltonians are optimal, as they are constructed to satisfy all the saturability conditions for the three-parameter bound in Eq.~\eqref{eqS:boundmagnetometryGS}.

To this purpose, we first need to find a GS satisfying
\begin{align*}
    \braket{J_x} = \braket{J_y} = \braket{J_z} = 0, \quad
    \mathrm{Var}_\rho(J_x) = \mathrm{Var}_\rho(J_y) = \mathrm{Var}_\rho(J_z) = \frac{N(N+2)}{12}.
\end{align*}
A pure state that satisfies these conditions, for $N=8n$ with $n \in \mathbb{N}$, is $\ket{\psi}$ defined in Eq.~\eqref{eq:3parGHZstate} (see Sec. \ref{app:variances} for an explicit calculation). It can be realized as the GS of
\begin{equation}
    H^C = a\ket{\psi}\bra{\psi} + H_\perp,
\end{equation}
where the support of $H_\perp$ is the $(\dim(\mathcal{H})-1)$-dimensional subspace orthogonal to $\ket{\psi}$, and all eigenvalues of $H_\perp$ are larger than $a$, with spectral gap $\Delta$.\\
Let us now focus on the first inequality in Eq.~\eqref{eq:inequalitiesGSmagnetometry}. To saturate it, we require a Hamiltonian such that $(\partial_k H_{\vec{\theta}})_{0i}=0$ for all $i\geq1$ with $E_i-E_0 \neq\Delta$ and for all $k=1, 2, 3$. Note that, in the non-degenerate case, satisfying this condition is generally not possible, as the derivatives do not commute. One can always choose the eigenvectors to satisfy the condition for a single derivative \cite{Abiuso_2025}, but it may then fail for the others.\\
The situation becomes simpler if degeneracy in the first excited state is allowed. For instance, the inequality can be saturated by choosing a Hamiltonian whose excited states are all degenerate, such as the one in Eq.~\eqref{eq:optimalGShamiltonian1}. However, constructing such a Hamiltonian is highly non-trivial, as it requires extremely precise tuning of the interactions between the particles.\\
To identify an optimal Hamiltonian with reduced degeneracy, we can exploit the following property of the state $\ket{\psi}$:
\begin{equation}
    \braket{\psi|J_i J_j|\psi} = 0 \quad \forall i\neq j,\, \forall N=4n,\, n\in \mathbb{N}.
\end{equation}
The proof, based on an explicit calculation, is provided in Sec.~\ref{app:covariancesGHZ}.\\
Let us now consider:
\[
   \ket{\psi_{i}} = \frac{J_i \ket{\psi}}{\sqrt{\braket{\psi |J_i^2| \psi}}}.
\]
We first note that these states are orthonormal, since $\braket{\psi|J_i J_k|\psi} = \delta_{ik}\braket{\psi |J_i^2| \psi}$. Moreover, $\braket{\psi|\psi_i}=\braket{\psi|J_i |\psi} = 0$, so $\{\ket{\psi}, \ket{\psi_1}, \ket{\psi_2}, \ket{\psi_3}\}$ form a set of orthonormal vectors. We complete the orthonormal basis by taking a set of orthonormal vectors $\{\ket{\tilde{\psi}_i}\}_{i=4}^{\dim(\mathcal{H})-1}$ that span the subspace orthogonal to the subspace generated by the other four vectors.\\
It is now clear that choosing the eigenvalues as in Eq.~\eqref{eq:optimalGShamiltonian2}, thereby reducing the degeneracy to only three states, allows us to saturate the first inequality in Eq.~\eqref{eq:inequalitiesGSmagnetometry}. Indeed, by definition,
\[
\braket{\psi |J_i|\tilde{\psi}_j}= \braket{\psi_i|\tilde{\psi}_j}=0.
\]
We are now left to check the diagonality condition and the WCC, which together require, for a pure state:
\begin{equation*}
    \braket{\partial_i \psi_{\vec{\theta}}|\partial_j \psi_{\vec{\theta}}}
    - \braket{\partial_i \psi_{\vec{\theta}}| \psi_{\vec{\theta}}}
      \braket{ \psi_{\vec{\theta}}|\partial_j \psi_{\vec{\theta}}}
    = 0 \quad \forall i \neq j.
\end{equation*}
From perturbation theory in quantum mechanics, we see that $\braket{\partial_i \psi_{\vec{\theta}}| \psi_{\vec{\theta}}}=0$, and therefore the condition reduces to
\begin{equation*}
    \braket{\partial_i\psi|\partial_j \psi} = 0 \quad \forall i \neq j.
\end{equation*}
We can compute the derivative as
\begin{equation}
    \ket{\partial_i \psi} = \sum_{k\geq 1} \frac{\braket{k|J_i|\psi}}{E_0-E_k} \ket{k},
\end{equation}
where for the first Hamiltonian $\ket{k} = \ket{\psi_{\perp,k}}$, while for the second one
\[
    \ket{k} =
    \begin{cases}
        \ket{\psi_k} & k=1,2,3, \\
        \ket{\tilde{\psi}_k} & k \geq 4.
    \end{cases}
\]
Thus, we can write $\braket{\partial_i\psi|\partial_j \psi}$ as
\[
    \braket{\partial_i\psi|\partial_j \psi}
    = \sum_{\alpha, \beta}
    \frac{\braket{0|J_i|\alpha}\braket{\beta|J_j|0}}{(E_0 - E_\alpha)(E_0 - E_\beta)}
    \braket{\alpha|\beta}
    = \sum_{\alpha} \frac{\braket{0|J_i|\alpha}\braket{\alpha|J_j|0}}{(E_0 - E_\alpha)^2}.
\]
In the fully degenerate case, this becomes:
\[
\braket{\partial_i\psi|\partial_j \psi}
= \frac{\sum_{\alpha}\braket{0|J_i|\alpha}\braket{\alpha|J_j|0}}{(E_0 - E_1)^2}
= \frac{\braket{0|J_i J_j |0}-\braket{0|J_i|0}\braket{0|J_j|0}}{\Delta^2}
= 0 \quad \forall i\neq j.
\]

In the other case, instead:
\begin{align*}
    \braket{\partial_i\psi|\partial_j \psi}
    = \frac{\sum_{k=1}^3\braket{0|J_i|\psi_k}\braket{\psi_k|J_j|0}}{\Delta^2}
    + \sum_{k=4}^{\dim(\mathcal{H})-1}
      \frac{\braket{0|J_i|\tilde{\psi}_k}\braket{\tilde{\psi}_k|J_j|0}}{(E_0 -E_k)^2}
    = 0 \quad \forall i\neq j.
\end{align*}
Indeed, $\braket{0|J_i|\tilde{\psi}_k}=0$, and
$\braket{0|J_i|\psi_k}\braket{\psi_k|J_j|0}\propto\delta_{ik}\delta_{jk}=0$ for
all $i \neq j$.\\
We thus have shown that both the diagonality condition and the WCC are satisfied. The corresponding QFIM is given by
\begin{equation*}
    Q = \frac{N(N+2)}{3\Delta^2}\,\mathbb{I}_3.
\end{equation*}

\subsubsection{Optimal measurement for three-parameter estimation with a ground state}

Finally, for both Hamiltonians, an optimal projective measurement composed of the projector onto the probe itself and projectors onto the orthogonal subspace, and which satisfies Eq.~\eqref{eq:purestateoptmeasorth}, is given by the projectors onto the following states:
\[
\ket{\omega_i}= \begin{cases}
    \ket{\psi} & i=0,\\
    \frac{\ket{\partial_i \psi}}{\sqrt{\braket{\partial_i \psi|\partial_i \psi}}} & i=1,2,3, \\
    \text{any orthonormal basis of the orthogonal subspace} & i\geq 4.
\end{cases}
\]

In particular, for the non-fully degenerate Hamiltonian, it is possible to evaluate the derivatives explicitly. Indeed, since $\braket{\tilde{\psi}_i | J_j | \psi} = \braket{\tilde{\psi}_i|\psi_j}=0$ for all $i$ and $j$, and $\braket{\psi_i|J_j|\psi} = \sqrt{\braket{\psi|J_j^2|\psi}} \delta_{ij}= \sqrt{\frac{N(N+2)}{12}} \delta_{ij}$, we obtain
\[
\ket{\partial_i \psi} = - \frac{1}{\Delta}\sqrt{\frac{N(N+2)}{12}}\ket{\psi_i}.
\]

Thus, in this case, we can choose the basis for the projective measurement to be the eigenbasis of the Hamiltonian itself:
\[
\ket{\omega_i}= \begin{cases}
    \ket{\psi} & i=0,\\
    \ket{\psi_i} & i=1,2,3, \\
    \ket{\tilde{\psi}_i} & i\geq 4.
\end{cases}
\]

\subsection{Optimal system for two-parameter estimation with a ground state}\label{app:optimalsystemtwoparGS}
Let us now focus on two-parameter GS magnetometry and consider the control Hamiltonian
\[
    H^C=g J_z^2 - J^2,
\]
so that the total Hamiltonian reads
\[
    H_{\vec{\theta}} = \theta_1 J_x + \theta_2 J_y + g J_z^2 - J^2. 
\]
The term $-J^2$ fixes the GS to the fully symmetric sector $j=N/2$, while the term $gJ_z^2$ lifts the degeneracy
within this sector. For even $N$ and $0<g\le N$, the ground state is the Dicke state $\ket{N/2,0}$, the first
excited level is spanned by $\ket{N/2,\pm1}$, at energy $g$ above the ground state, and the gap is $\Delta=g$
(the lowest state with $j=N/2-1$ lies at energy $N$ above the ground state). Since
\begin{align*}
  J_x\ket{\tfrac N2,0}&=\tfrac12\sqrt{\tfrac N2\big(\tfrac N2+1\big)}\big(\ket{\tfrac N2,1}+\ket{\tfrac N2,-1}\big),\\
  J_y\ket{\tfrac N2,0}&=\tfrac{1}{2i}\sqrt{\tfrac N2\big(\tfrac N2+1\big)}\big(\ket{\tfrac N2,1}-\ket{\tfrac N2,-1}\big),
\end{align*}
the derivatives of the Hamiltonian couple the ground state only to the first excited level, and
Eq.~\eqref{eqS:QFIMGS} gives directly
\begin{equation*}
  Q_{11}=Q_{22}=\frac{4\,\mathrm{Var}_{\ket{\frac N2,0}}(J_x)}{g^2}=\frac{N(N+2)}{2g^2},\qquad
  Q_{12}=\frac{4\,\mathrm{Re}\braket{\tfrac N2,0|J_xJ_y|\tfrac N2,0}}{g^2}=0 .
\end{equation*}
Hence $Q=\frac{4}{\Delta^2}\Gamma$ with a diagonal $\Gamma$, and the variances
$\mathrm{Var}(J_x)=\mathrm{Var}(J_y)=N(N+2)/8$ satisfy Eq.~\eqref{eq:optimalvariances}, so that the $d=2$ bound is
saturated. Finally, since $J_x\ket{N/2,0}$ and $J_y\ket{N/2,0}$ lie in the first excited level, first-order
perturbation theory gives $\ket{\partial_1\psi}=-J_x\ket{\tfrac N2,0}/g$ and $\ket{\partial_2\psi}=-J_y\ket{\tfrac N2,0}/g$,
so that $\braket{\partial_1\psi|\partial_2\psi}=\braket{\tfrac N2,0|J_xJ_y|\tfrac N2,0}/g^2=0$ and the
WCC~\eqref{eq:WCCpurestate} is satisfied.

\subsubsection{Optimal measurement for two-parameter estimation with a ground state}

Finally, an optimal projective measurement composed of the projector onto the probe itself and projectors onto the orthogonal subspace, and which satisfies Eq.~\eqref{eq:purestateoptmeasorth}, is given by
\[
\Big\{\ket{0} \coloneqq \ket{\tfrac{N}{2}, 0}, \;
\ket{\pm1} \coloneqq \frac{\ket{\tfrac{N}{2}, 1} \pm \ket{\tfrac{N}{2}, -1}}{\sqrt{2}} \Big\},
\]
together with any orthonormal basis for the orthogonal subspace. Note that $\ket{\pm1}$ are respectively proportional to $\ket{\partial_1\psi_{\vec{\theta}}}$ and $\ket{\partial_2\psi_{\vec{\theta}}}$.

\section{Covariances of the angular momenta over the XYZ-GHZ state}
\subsection{Variances of the angular momentum}\label{app:variances}
\paragraph*{Proposition 2.} For $N=8n$, the XYZ-GHZ state $\ket{\psi}$ of Eq.~\eqref{eq:3parGHZstate} satisfies
\[
  \braket{J_i}=0,\qquad \frac12\braket{\{J_i,J_j\}}=\frac{N(N+2)}{12}\,\delta_{ij}.
\]
For $N=4n$, $\braket{J_i}=0$ and $\braket{J_iJ_j}=0$ for $i\neq j$ still hold (Sec.~\ref{app:covariancesGHZ}),
while for odd $n$ the variances deviate from $N(N+2)/12$ by relative corrections of order $2^{-N/2}$.
We first show that the first moments vanish for $N=4n$.

To compute the variances, we need:
\begin{align*}
    \braket{-|0} &= \braket{+|0} =\braket{+|1}= \frac{1}{\sqrt{2}},\\
    \braket{-|1} &=- \frac{1}{\sqrt{2}},\\
    \braket{0|\pm y}&= \frac{1}{\sqrt{2}},\\
    \braket{1|\pm y}&= \frac{\pm i}{\sqrt{2}},\\
    \braket{+|\pm y}&= \frac{1\pm i}{2},\\
    \braket{-|\pm y}&= \frac{1\mp i}{2}.
\end{align*}
Let us first compute
\[
    \braket{\psi|J_z|\psi}=\frac{\mathcal{N}^2}{2}\Big(\sum_{i=1}^3 (\bra{\phi_i^+}^{\otimes N} + \bra{\phi_i^-}^{\otimes N})\Big) J_z \Big(\sum_{i=1}^3 (\ket{\phi_i^+}^{\otimes N} + \ket{\phi_i^-}^{\otimes N}) \Big).
\]
The scalar product contains the following contributions:
\begin{enumerate}
    \item  The following terms vanish since $J_z$ creates a superposition of states with 1 spin flipped, all orthogonal to the original state:
    \begin{align*}
        \braket{\phi_1^\pm |^{\otimes N} J_z |\phi_1^\pm}^{\otimes N} &= \braket{\phi_1^\pm |^{\otimes N} J_z |\phi_1^\mp}^{\otimes N} = \braket{\phi_2^\pm |^{\otimes N} J_z |\phi_2^\pm}^{\otimes N} = \braket{\phi_2^\pm |^{\otimes N} J_z |\phi_2^\mp}^{\otimes N}\\
        &= 0.
    \end{align*}
        
    \item  The two terms always cancel out: \[
    \braket{\phi_3^\pm |^{\otimes N} J_z |\phi_3^\pm}^{\otimes N}= \bra{\phi_3^\pm   }^{\otimes N}\sum_{i=1}^N\frac{\sigma_3^{(i)}}{2}\ket{\phi_3^\pm}^{\otimes N}=\pm \frac{N}{2}.
    \]
    \item The following two terms vanish:
    \[
    \braket{\phi_3^\pm |^{\otimes N} J_z |\phi_3^\mp}^{\otimes N}=0.
    \]
    \item The following four terms cancel out if $N$ is even:
    \begin{align*}
        \braket{\phi_3^+ |^{\otimes N} J_z |\phi_1^\pm}^{\otimes N} &= \frac{N}{2}(\braket{0|\pm})^{N-1}\braket{0|\mp} = \frac{N}{ 2^{\frac{N}{2}+1}}, \\
        \braket{\phi_3^- |^{\otimes N} J_z |\phi_1^\pm}^{\otimes N}&= -\frac{N}{ 2^{\frac{N}{2}+1}}.
    \end{align*}
    \item The following eight terms cancel if $N=4n$:
\begin{align*}
    \braket{\phi_3^- |^{\otimes N} J_z |\phi_2^\pm}^{\otimes N} &= \frac{N}{2} (\braket{1|\pm y})^{N-1}\braket{1|\mp y} = \frac{N}{2} \Big(\pm \frac{i}{\sqrt{2}}\Big)^{4n-1}\Big(\mp \frac{i}{\sqrt{2}}\Big)
    =\frac{N}{2^{\frac{N}{2}+1}} (\pm i)^{-1} (\mp i) = - \frac{N}{ 2^{\frac{N}{2}+1}}\\
    &= - \braket{\phi_3^+ |^{\otimes N} J_z |\phi_2^\pm}^{\otimes N},\\
    \braket{\phi_1^+ |^{\otimes N} J_z |\phi_2^\pm}^{\otimes N} &= \frac{N}{2^{N+1}}(1 \pm i)^{4n-1}(1 \mp i)  = \frac{\mp iN(-4)^n}{ 2^{N+1}}
    = - \braket{\phi_1^- |^{\otimes N} J_z |\phi_2^\pm}^{\otimes N}.
    \end{align*}
    \item Consequently, the contributions from the complex conjugate of the previous terms also cancel.
\end{enumerate}

Thus, to achieve $\braket{J_z}=0$, it is sufficient to take $N=4n$. For $N \neq 4 n$, corrections are exponentially small in $N$. By symmetry, $\braket{J_x} = \braket{J_y} = 0$.\\

For the second moments we use symmetry. The single-qubit unitary $C=[\mathbb{I}-i(\sigma_x+\sigma_y+\sigma_z)]/2$,
a rotation by $2\pi/3$ about the axis $(1,1,1)$, satisfies $C\sigma_zC^\dagger=\sigma_x$, $C\sigma_xC^\dagger=\sigma_y$
and $C\sigma_yC^\dagger=\sigma_z$, and maps $\ket{\phi_3^\pm}\to\ket{\phi_1^\pm}\to\ket{\phi_2^\pm}\to\ket{\phi_3^\pm}$
up to phases that are integer powers of $e^{i\pi/4}$. Hence, for $N=8n$, $C^{\otimes N}$ maps
$\ket{\Phi_3}\to\ket{\Phi_1}\to\ket{\Phi_2}\to\ket{\Phi_3}$, so that $C^{\otimes N}\ket{\psi}=\ket{\psi}$, while
$C^{\otimes N}J_zC^{\dagger\otimes N}=J_x$ and $C^{\otimes N}J_xC^{\dagger\otimes N}=J_y$. Therefore
$\braket{\psi|J_x^2|\psi}=\braket{\psi|J_y^2|\psi}=\braket{\psi|J_z^2|\psi}$. Since $\ket{\psi}$ lies in the
symmetric subspace, where $\vec J^2=\frac N2(\frac N2+1)$, and $\braket{J_i}=0$,
\[
  \mathrm{Var}_{\ket{\psi}}(J_i)=\frac13\braket{\psi|\vec J^2|\psi}=\frac{N(N+2)}{12},\qquad i=x,y,z .
\]
For $N=4n$ with $n$ odd, the phases above do not cancel, and the three second moments differ from
$N(N+2)/12$ by relative corrections of order $2^{-N/2}$.

\subsection{Covariances of the angular momenta}\label{app:covariancesGHZ}

Here we prove that
\begin{equation*}
    \mathrm{Cov}_{\ket{\psi}}(J_i, J_j) = 0 \quad \forall\, i\neq j,\;
    \forall\, N=4n,\; n\in \mathbb{N}.
\end{equation*}
Since we already proved that $\braket{J_i}=0$, it is enough to prove that
$\braket{\psi|J_i J_j|\psi} = 0$ for all $i \neq j$.

Let us first consider $\braket{\psi|J_x J_y|\psi}$. For the calculation we
need
\begin{align*}
    \sigma_x \ket{0} &= \ket{1}, & \sigma_y \ket{0} &= i\ket{1},\\
    \sigma_x \ket{1} &= \ket{0}, & \sigma_y \ket{1} &= -i\ket{0},\\
    \sigma_x \ket{\pm y} &= \pm i \ket{\mp y}, &
    \sigma_y \ket{\pm} &= \mp i\ket{\mp}.
\end{align*}
We now explicitly calculate the contributions to
$\braket{\psi|J_x J_y|\psi}$:
\begin{align*}
    \bra{0}^{\otimes N} J_x J_y \ket{0}^{\otimes N}
      &= \frac{Ni}{4} = - \bra{1}^{\otimes N} J_x J_y \ket{1}^{\otimes N}, \\
    \bra{\pm}^{\otimes N} J_x J_y \ket{\pm}^{\otimes N}
      &= \bra{\pm y}^{\otimes N} J_x J_y \ket{\pm y}^{\otimes N}
       = \bra{0}^{\otimes N} J_x J_y \ket{1}^{\otimes N}
       = \bra{1}^{\otimes N} J_x J_y \ket{0}^{\otimes N} \\
      &= \bra{\pm y}^{\otimes N} J_x J_y \ket{\mp y}^{\otimes N}
       = \bra{\pm}^{\otimes N} J_x J_y \ket{\mp}^{\otimes N} = 0,\\
    \bra{0}^{\otimes N} J_x J_y \ket{\pm}^{\otimes N}
      &= -\frac{i N(N-2)}{4\cdot 2^{N/2}}
       = - \bra{1}^{\otimes N} J_x J_y \ket{\pm}^{\otimes N},\\
    \bra{\pm}^{\otimes N} J_x J_y \ket{0}^{\otimes N}
      &= \frac{i N^2}{4\cdot 2^{N/2}}
       = - \bra{\pm}^{\otimes N} J_x J_y \ket{1}^{\otimes N},\\
    \bra{0}^{\otimes N} J_x J_y \ket{\pm y}^{\otimes N}
      &= \frac{i N^2}{4\cdot 2^{N/2}}
       = - \bra{1}^{\otimes N} J_x J_y \ket{\pm y}^{\otimes N},\\
    \bra{\pm y}^{\otimes N} J_x J_y \ket{0}^{\otimes N}
      &= -\frac{i N(N-2)}{4\cdot 2^{N/2}}
       = - \bra{\pm y}^{\otimes N} J_x J_y \ket{1}^{\otimes N},\\
    \bra{\pm}^{\otimes N} J_x J_y \ket{+y}^{\otimes N}
      &= \pm \frac{N^2 (-1)^n}{4 \cdot 2^{N/2}}
       = - \bra{\pm}^{\otimes N} J_x J_y \ket{-y}^{\otimes N},\\
    \bra{+y}^{\otimes N} J_x J_y \ket{\pm}^{\otimes N}
      &= \pm \frac{N(N-2) (-1)^n}{4 \cdot 2^{N/2}}
       = - \bra{-y}^{\otimes N} J_x J_y \ket{\pm}^{\otimes N}.
\end{align*}
All contributions cancel, hence $\braket{\psi|J_x J_y|\psi}=0$. By symmetry,
$\braket{\psi|J_i J_j|\psi} = 0$ for all $i \neq j$ and all $N = 4n$.

\section{Rank optimization and worst-case sensitivity}
\label{app:RANK_MAX}

There are other figures of merit, beyond the weighted mean square error $\mathrm{Tr}[WQ^{-1}]$, that can be meaningful to multiparameter metrology in different operational contexts. One example is the Bures volume of sensitivity (or inverse sloppiness, $\det Q$).
For example, in the recent~\cite{he2026fishergeometry} even the tradeoff between sloppiness and multiparameter incompatibility is structurally analyzed. In yet another direction, the authors of~\cite{oconnor2026geometricobstructions} analyze how many independent parameters can be estimated with a Heisenberg-like $t^2$ scaling in dynamical metrology, as opposed to the constant $t^0$ sensitivity. A similar split can be found for the $\beta^2$ vs $\Delta^{-2}$ scaling in equilibrium metrology when analyzing the diagonal and off-diagonal contributions to the QFI.

In this section, we focus on a different figure of merit for multiparameter metrology, namely we ask \emph{how many independent parameters a probe can be sensitive to}~\cite{Candeloro_2024}.
As we will see, this showcases stark differences among the three main encodings considered in this work (dynamical, thermal, ground state). Notice also that once the number of parameters exceeds the maximum rank of $Q$, the associated Bures volume of sensitivity $\det Q$ necessarily becomes null, and as such the sloppiness infinite.

{\bf Rank minimization.} Notice also that in the context of distributed private quantum sensing, one actually tries to \emph{minimize} the rank of $Q$, so to make it susceptible to a single linear combination of the parameters involved, while being transparent to the other linearly independent combinations~\cite{hassani2025privacy}. Given the results of this section, full-rank thermal states are not naturally suited to private distributed sensing when the Hamiltonian can vary along a complete basis of traceless directions, whereas ground states (as well as dynamically evolving states) can be. The argument is as follows: if a thermal state $\rho$ has all populations different from zero, then the positivity of the QFI kernel, expressed in the diagonal basis of $\rho$ $\frac{2(p_i-p_j)^2}{(\ln p_i - \ln p_j)^2 (p_i + p_j)}$ implies that the rank will actually \emph{always be} $D^2-1$. The only option to reduce the rank is to collapse some of the populations. In the limit of collapsing all-but-one, one is left with the ground-state problem.

{\bf Rank maximization.}
Consider a quantum system of Hilbert space dimension $D$. Dynamical, thermal, and ground-state Hamiltonian encodings are all insensitive to global shift of the energy levels, i.e. no transformation of the form $H\rightarrow H+\alpha\mathbf{1}$ can be detected.
As such, we consider a complete orthonormal basis of the physically-distinguishable Hamiltonians, namely we consider the equivalence classes generated by the equivalence relation 
\begin{align}
    H_1\simeq H_2 \quad \text{if}\quad H_1-H_2\propto\mathbf{1}\;.
\end{align} 
The resulting set can be then taken as the traceless hermitian operators, with a Hilbert-Schmidt orthonormal basis
\begin{align}
    \Tr[h_i h_j]=\delta_{ij}\;, \quad \Tr[h_i]=0\;, \quad h_i^\dagger=h_i\quad i,j=1,\dots, D^2-1\;.
\end{align}
These operators form a $D^2-1$ dimensional real vector space
\begin{align}
    \mathcal{H}:={\sf Span}\{h_i\} \cong \mathbb{R}^{D^2-1} \;,
\end{align}
which we use to fully parametrize all possible parameter encodings locally (i.e., local Hamiltonian variations) as
\begin{align}
\label{eqapp:thetaD21basis}
    H_{\vec{\theta}} = H^{P}_0+\vec{\theta}\cdot\vec{h}+H^C\;.
\end{align}
We want to characterize the amount of independent parameters to which our probes can be, in principle, sensitive. For that we notice 
\begin{align}
    {\sf rank}[Q]={\sf dim}\;\mathcal{H} - {\sf dim}\; V_\parallel = D^2-1 - {\sf dim} V_\parallel\;,
\end{align}
where $V_\parallel$ is the subspace of $\mathbb{R}^{D^2-1}$ that does not generate any variation in $\rho_{\vec{\theta}}$, also corresponding to the kernel of $Q$
\begin{align}
    {\sf ker} \; Q \equiv V_\parallel:=\left\{\vec{\theta}\;\bigg|\; \sum_i \theta_i \partial_i\rho =0 \right\}\;.
\end{align}
As we will see, this point of view marks a stark difference among the encodings we consider: out of the $D^2-1$ directions in which a Hamiltonian can be perturbed, only thermal states can be sensitive to all, whereas ground states and pure unitarily evolving states are bound to be sensitive at most to $2(D-1)$ parameters. 
The case of unitarily evolving non-pure states is slightly more complicated, and it depends on the eigenvalue degeneracies.

We summarize the main bounds on the rank of the QFIM in Table~\ref{tab:Q_ranks}, and derive them in the subsections below, case by case.

\begin{table}[htbp]
\centering
\begin{tabular}{c|c}
\hline
\textbf{Encoding} & \textbf{Maximum rank of} $Q_{ij}$  \\ \hline \\[-4pt]

Thermal: 
$\rho_{\vec{\theta}} = \frac{e^{-\beta H_{\vec{\theta}}}}{\operatorname{Tr}(e^{-\beta H_{\vec{\theta}}})}$
& $D^2-1$   \\[12pt] \hline \\[-4pt]

GS of $H_{\vec{\theta}}$  
with spectral gap $\Delta$ & $2(D-1)$  \\[12pt] \hline \\[-4pt]

Dynamical pure:
$\rho_{\vec{\theta}} = e^{-i H_{\vec{\theta}} t / \hbar} \ketbra{\psi}\, e^{i H_{\vec{\theta}} t / \hbar}$
& $2(D-1)$  \\[12pt] \hline \\[-4pt]

Dynamical general:
$\rho_{\vec{\theta}} = e^{-i H_{\vec{\theta}} t / \hbar} \rho_0\, e^{i H_{\vec{\theta}} t / \hbar}$
&  $\sum_{\alpha\neq \beta}  g_\alpha g_\beta \leq D(D-1)$  \\[12pt] \hline 
\end{tabular}
\caption{When considering quantum systems of Hilbert space dimension $D$, the amount of linearly independent parameters encodable in a Hamiltonian is $D^2$. However, scalar shifts $H\rightarrow H+\alpha\mathbf{1} $ do not lead to any observable variation in thermal states, ground states, or unitary dynamics. As such, we consider a basis of $D^2-1$ traceless hermitian Hamiltonian perturbations, and study the maximum number of independent parameters to which each encoding can be sensitive. This corresponds to the rank of the quantum Fisher information matrix $Q$. We find that in general only thermal states can achieve a complete ${\sf rank}\;Q=D^2-1$, whereas pure states (both ground states and dynamically evolved pure states) can only be sensitive to $2(D-1)$ parameters. For the case of generic unitarily evolving mixed states, the bound depends on the values $g_\alpha$ of all degeneracies of the eigenvalues of $\rho_0$ (one recovers the result for pure states $g_1=1, g_0=D-1$, for completely mixed states $g_{1/D}=D$ a null rank, whereas the maximum possible rank is obtained for mixed states with no degeneracy in which $g_\alpha=1\; \forall\alpha$).}
\label{tab:Q_ranks}
\end{table}

\subsection{Thermal case}
\label{app:rank_therm}
The estimation of the maximum rank of $Q$ in the thermal case is simple. It is sufficient to consider
\begin{align}
    H_0^{P}+H^C=0 \quad \text{in order to obtain}\quad \rho_{\vec{\theta}}=\frac{e^{-\beta \vec{\theta}\cdot \vec{h}}}{Z}\;,
\end{align}
that is $\rho_0\equiv \frac{\mathbf{1}}{D}$ and any nonzero $\vec{\theta}$ induces a linear variation to the state. As such the kernel $V_\parallel$ is trivial, $V_\parallel=\{0\}$, and the QFIM $Q$ is full rank. More precisely, from Eq.~\eqref{eqS:QFIMthermal} applied to $\rho=\frac{\mathbf{1}}{D}$ we get
\begin{align}
    Q_{ij} = \frac{\beta^2}{D} \mathrm{Tr}[h_i h_j]=\frac{\beta^2}{D}\delta_{ij} .
\end{align}

\subsection{Ground state case}
\label{app:rank_ground}
One has for ground states
\begin{align}
    Q_{ij} = \sum_{k\neq 0} \frac{4\,\mathrm{Re}[\partial_i H_{0k} \partial_j \bar{H}_{0k}]}{(E_k - E_0)^2} + \mathcal{O}(\beta^2e^{-\beta\Delta}),
\end{align}
and around the base point
\begin{align}
    H_0^P+H^C=\sum_{k}E_k\ketbra{e_k}{e_k} \quad \text{and} \quad H_{\vec{\theta}}=H_0^P+H^C+\vec{\theta}\cdot \vec{h}\;.
\end{align}
we choose then a basis including the $2(D-1)$ elements
\begin{align}
    h_{2i-1}=\frac{\ketbra{e_i}{e_0}+\ketbra{e_0}{e_i}}{\sqrt{2}}\;, \quad h_{2i}=\frac{\ketbra{e_i}{e_0}-\ketbra{e_0}{e_i}}{i\sqrt{2}} \quad i=1,\dots,D-1
\end{align}
such that the variation in the ground state induced by any further $h_{i>2(D-1)}$ is null, as all other basis elements will commute with $\ketbra{e_0}$.
We therefore see that for ground state encoding we have ${\sf rank}\;Q=D^2-1-{\sf dim}V_\parallel=2(D-1)$.

\subsection{Dynamical case}
\label{app:rank_dyn}
Given an evolution of the form
\begin{align}
    \rho_\theta =U_\theta\rho U^\dagger_\theta := e^{-i H_\theta \frac{t}{\hbar}}\rho  e^{i H_\theta \frac{t}{\hbar}}
\end{align}
one has
\begin{align}
    \partial_i \rho_\theta = -i \frac{t}{\hbar}U_\theta[\tilde{h}_i,\rho]U_\theta^\dagger \quad \text{where}\quad \tilde{h}_i=\int_0^1 ds\, e^{is H_{\vec{\theta}}\frac{t}{\hbar}} (\partial_i H_\theta) e^{-is H_{\vec{\theta}}\frac{t}{\hbar}}=\int_0^1 ds\,U_\theta^{-s} h_i U_\theta^s \;,
\end{align}
as again we assume the full $D^2-1$ dimensional basis of parameters~\eqref{eqapp:thetaD21basis}.
Now consider the map
\begin{align}
    T[h]=\int_0^1 ds\;U_\theta^{-s} h U_\theta^s \;.
\end{align}
We thus want to bound the dimension of the parameter subspace that induces a nonzero variation in $\rho_\theta$, i.e. $\vec{\theta}\cdot[T[\vec{h}],\rho]\neq 0$. Formally we define the two subspaces
\begin{align}
    V'_\parallel: & \left\{ h\bigg| [h,\rho]=0\right\}\;.\\
    V_\parallel: & \left\{ h\bigg| [T[h],\rho]=0\right\}\;.
\end{align}
We are interested in ${\sf dim} V_\parallel$. To bound it, consider first $\rho_\theta$ (or equivalently $\rho$) having eigenvalues $\lambda_\alpha$ with degeneracies $g_\alpha$ such that $\sum_\alpha g_\alpha=D$. It is then easy to verify
\begin{align}
    {\sf dim} V'_\parallel=D^2-1 - \sum_{\alpha\neq \beta} g_\alpha g_\beta \;,
\end{align}
 that is the dimension of the commutator space of $\rho_\theta$.
To bound ${\sf dim}V_\parallel$, we then consider
\begin{align}
    {\sf dim}V_\parallel &= {\sf dim} \ker [T] +  {\sf dim} ({\rm Im} T \cap V'_\parallel)        \\
    & \geq       {\sf dim} \ker [T] +  {\sf dim} ({\rm Im} T) +{\sf dim}  V'_\parallel - {\sf dim} \mathcal{H} \\
    & = {\sf dim}  V'_\parallel \;.
\end{align}
As such we find $ {\sf dim}V_\parallel\geq D^2-1 - \sum_{\alpha\neq \beta} g_\alpha g_\beta$ and
\begin{align}
    {\sf rank }\; Q=D^2-1-{\sf dim} V_\parallel \leq \sum_{\alpha\neq \beta} g_\alpha g_\beta
\end{align}
completing Table~\ref{tab:Q_ranks} in the general unitary case.

\subsection{Tightness of the bounds}
Are the bounds on the rank of $Q$ derived above and summarized in Table~\ref{tab:Q_ranks} tight?

The answer in the thermal case is trivially yes, as the construction only involves preparing the maximally mixed state as the unperturbed thermal state, regardless of the choice of $h_i$ in which the parameters are encoded.

The case of ground state encoding and dynamical state encoding are more subtle, and the answer is in general negative.
For both a pure ground state and a pure dynamical state, the question of whether they can be sensitive to a perturbation induced by $h_i$ reduces to \emph{whether the state commutes} with $h_i$ (or $\tilde{h}_i$ in the dynamical case, see Sec.~\ref{app:rank_dyn}).
In other words, in order to understand whether the {\sf rank}$Q\leq 2(D-1)$ is tight, we ask: given $2(D-1)$ Hilbert-Schmidt orthonormal hermitian operators, is it always possible to find a pure state that does not commute with any of them (nor linear combinations of them)? 
The answer is not always. For example, consider $D=3$ and the 4-dimensional space of traceless hermitian operators of the form
\begin{align}
    \begin{pmatrix}
        a& b & 0 \\
        \bar{b} & c & 0 \\
        0 & 0 & -(a+c)
    \end{pmatrix}\;.
\end{align}
Any vector in $\mathbb{C}^3$ can be an eigenvector of such matrix for an appropriate choice of $a,b,c$. This means that the upper bound to {\sf rank}$Q$ for the ground state case (as well as for the pure dynamical case) is not always tight.

\end{document}